\documentclass[twocolumn,secnumarabic,amssymb, nobibnotes, aps, prd]{revtex4-2}

\usepackage{natbib}
\usepackage{amsmath}
\usepackage{graphicx}
\usepackage{xcolor}

\begin{document}
\title{Absence of sizable superconductivity in hydrogen boride: A first principles study}
\author{Antonella Meninno$^{1,2}$}\email[]{ameninno001@ikasle.ehu.es}\author{ Ion Errea$^{1,2,3}$}

\affiliation {$^1$Centro de F\'isica de Materiales (CSIC-UPV/EHU), Manuel de Lardizabal Pasealekua 5, 20018 Donostia/San Sebasti\'an, Spain}

\affiliation {$^2$Fisika Aplikatua 1 Saila, Gipuzkoako Ingeniaritza Eskola, University of the Basque Country (UPV/EHU), Europa Plaza 1, 20018 Donostia/San Sebasti\'an, Spain} 

\affiliation {$^3$Donostia International Physics Center (DIPC), Manuel de Lardizabal Pasealekua 4, 20018 Donostia/San Sebasti\'an, Spain}

\begin{abstract}
The recently synthesized hydrogen boride monolayer in the $Cmmm$ phase is a promising superconductor due to its similarity to MgB$_2$ and the large hydrogen content in its structure. Making use of first-principles calculations based on density functional theory, we study its electronic, vibrational, and superconducting properties and conclude that, despite the expectations, hydrogen boride does not have a sizable superconducting critical temperature. The presence of hydrogen in the system alters the boron-boron bonding, weakening the electron-phonon interaction. We have studied the effect of enhancing the critical temperature by doping the system, but the inclusion of electrons or holes reveals ineffective. We attribute the small critical temperature of this system to the vanishing hydrogen character of the states at the Fermi level, which are dominated by boron $p$ states. Our results determine that a large proportion of hydrogen-like states are needed at the Fermi level to attain a large superconducting critical temperature in hydrogenated monolayers.
\end{abstract}

\maketitle

\section{Introduction}

\label{intro}

The observation of high temperature superconductivity in hydrogen-based superconductors, with critical temperatures ($T_c$'s) surpassing 200 K, is one of the most astonishing results in physics of the last years. Some examples of such  ``superhydrides'' are H$_3$S \cite{drozdov2015conventional}, with a maximum $T_c$ of 203 K at 155 GPa; LaH$_{10}$ \cite{somayazulu2019evidence,drozdov2019superconductivity}, with $T_c = 250$ K at 150 GPa; YH$_9$ \cite{snider2021synthesis,kong2021superconductivity}, reaching a $T_c$ of around 250 K at approximately 200 GPa; and YH$_6$ \cite{kong2021superconductivity,semenok2020superconductivity}, with $T_c=224$ K at 166 GPa. 

The actual challenge is to reduce the pressure at which superconductivity occurs in superhydrides, with the ultimate goal of understanding if high-$T_c$ superconductivity is possible at ambient pressure for these type of compounds. Interestingly, the enhancement of superconductivity by hydrogen absorption is a well known phenomenon, observed for example in palladium and thorium hydrides at ambient pressure \cite{skoskiewicz1972superconductivity,satterthwaite1970superconductivity}. In both cases, the superconducting critical temperature of the original element (Pd or Th) increases from around 1 K up to approximately 10 K in the stoichiometric hydride. The role of hydrogen in enhancing the $T_c$ has been further studied in the context of monolayers at ambient pressures. In previous theoretical works based on first principles calculations, it has been shown that a MgB$_2$ monolayer, which has a critical temperature of 39 K in the bulk \cite{nagamatsu_superconductivity_2001}, can reach a $T_c$ of 69 K after hydrogenation \cite{bekaert2019hydrogen}, and that $p$-doped graphane, a hydrogenated graphene monolayer, reaches a critical temperature of around $80$ K even if graphene itself is not superconducting \cite{savini2010first}. Hydrogenated monolayers offer thus a promising platform to synthesize high-$T_c$ hydrogen-based compounds at ambient pressures, overcoming the limits of the high-pressure superhydrides.

\begin{figure*}
\begin{minipage}{0.49\textwidth}

\centering
\includegraphics[width=\columnwidth]{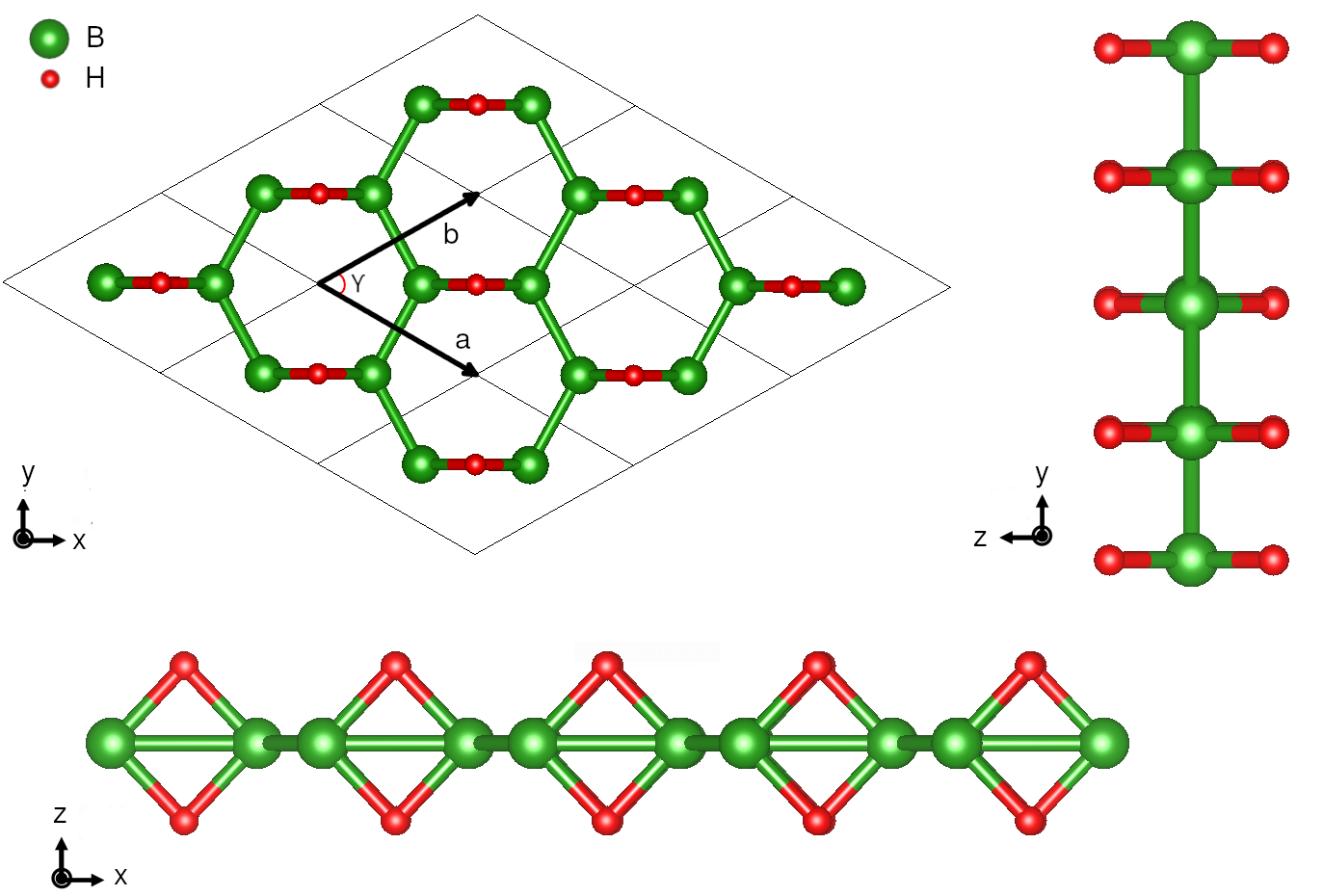}
\caption{Different views of the crystal structure of the hydrogen boride (HB) monolayer in the $Cmmm$ phase. The lattice parameters are indicated, together with the $\gamma$ angle}
\label{structure}

\end{minipage}\hfill
\begin{minipage}{0.49\textwidth}

\centering
\includegraphics[width=\columnwidth]{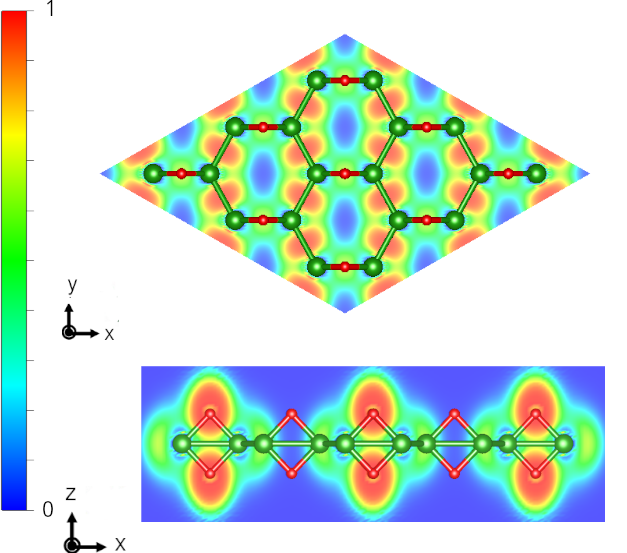}
\caption{Electron localization function (ELF) for the non doped HB monolayer. In the top panel the $z=0$ plane is represented, while in the bottom panel the $y=0$ plane (putting the origin of coordinates at the position where the lattice vectors in Fig. \ref{structure} start).}
\label{elfnodop}

\end{minipage}
\end{figure*}

The recently synthesized stoichiometric hydrogen boride monolayer (HB) \cite{nishino2017formation} is a promising superconductor \cite{tateishi2019semimetallicity}. HB was obtained through a procedure of wet chemical exfoliation at ambient pressure from MgB$_2$ through ion-exchange treatment, and it is believed to adopt a quasi-hexagonal structure with a boron honeycomb layer reminiscent of the structure of MgB$_2$, with H atoms out of plane (see Fig. \ref{structure}) \cite{nishino2017formation,kawamura2019photoinduced,tateishi2019semimetallicity}. The structure of HB observed experimentally seems to adopt a $Cmmm$ space group as anticipated by \emph{ab initio} crystal structure prediction calculations \cite{jiao2016two-dimensional}. More recently, different polymorphs of hydrogen and boron have been synthesized in the two-dimensional limit \cite{li2021synthesis}, showing the potential of ``borophane'' to form different stable compounds at the nanoscale and broaden its applications in nanodevices. 

Despite its semimetallic character \cite{tateishi2019semimetallicity}, $Cmmm$ HB is a good superconducting candidate for several reasons. Mainly because several parent ``borophene'' monolayers have been predicted to have $T_c$'s of around 20 K \cite{ga02017prediction} and hydrogenation, as mentioned above, is expected to enhance it. Also, because the bonding nature of boron in $Cmmm$ HB is reminiscent of the $sp^2$ bonding in MgB$_2$, which is crucial for its large $T_c$, even if hydrogen has a large impact on the band structure of the compound \cite{tateishi2019semimetallicity}. The goal of the present work is to determine the potential superconductivity of $Cmmm$ HB through first principles calculations based on density-functional theory (DFT). We have also analyzed the possibility of enhancing the critical temperature of the system by doping.

The paper is organized as follows. In section \ref{computational} we overview the theoretical framework of the calculations performed, in section \ref{results} we present the results of our calculations, in section \ref{discussion} we analyze and interpret the results further, and in section \ref{conclusion} we summarize the results.

\section{Computational details}

\begin{figure*}
\centering
\includegraphics[width=\textwidth]{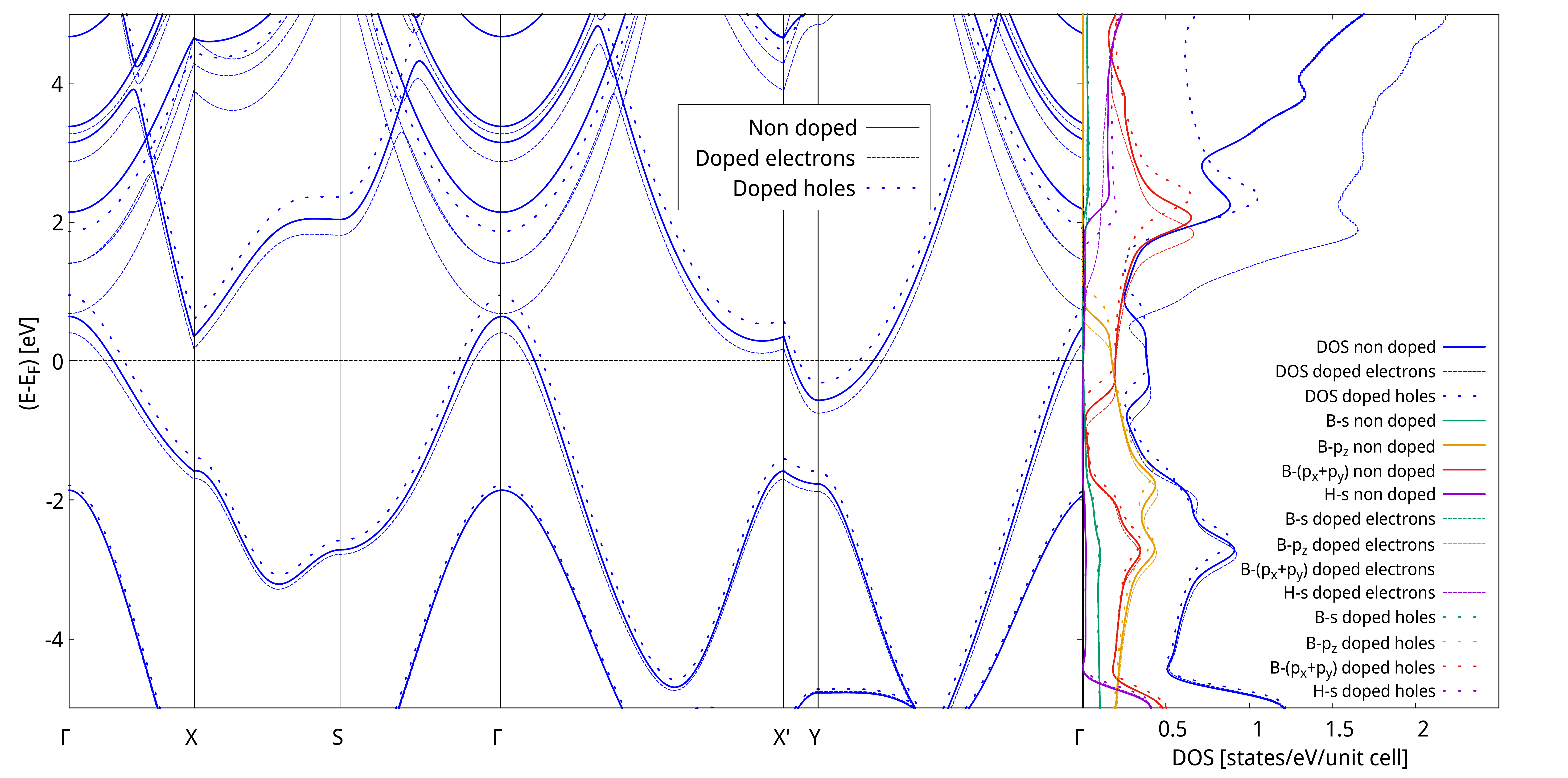}
\caption{Electronic band structure (left panel) and density of states (DOS) (right panel) for $Cmmm$ hydrogen boride. The bands and DOS are also plotted with electron and hole doping. The projection of the DOS onto atomic orbitals is also included.}
\label{bandsanddos}
\end{figure*}

\label{computational}
 
All computations have been performed using DFT within the {\sc Quantum ESPRESSO} package \cite{giannozzi2009quantum,giannozzi2017advanced}, making use of the Perdew-Burke-Ernzerhof (PBE) parametrization of the exchange-correlation functional \cite{perdew1996generalized}. The electron-ion interaction has been modeled making use of ultrasoft pseudopotentials, including $2s^2$ and $2p^1$ electrons of boron in the valence. The self-consistent DFT computation has been performed with a grid of 42$\times$42$\times$1 {\bf k} points for the integrals over the Brillouin zone and a Methfessel-Paxton  first-order spreading smearing of 0.02 Ry \cite{PhysRevB.40.3616}. A kinetic energy cutoff of $50$ Ry has been used for the plane-wave basis and a $500$ Ry cutoff for the charge density. In order to avoid spurious interactions, we have left a vacuum between monolayer replicas of 35 $a_0$. Phonon frequencies have been computed in the harmonic approximation within density-functional perturbation theory (DFPT) \cite{baroni1987green,baroni2001phonons}. Dynamical matrices have been calculated explicitly in an 8$\times$8$\times$1 grid of {\bf q} points, and we have used Fourier interpolation to obtain the phonon spectrum. 

The electron-phonon interaction has been computed within DFPT. The electron-phonon contribution to the phonon linewidth \cite{allen1972neutron} can be approximated at low temperatures as 
\begin{eqnarray}
\label{linewidth}
\gamma_\nu({\bf q}) & = &\frac{2\pi\omega_\nu({\bf q})}{N_{{\bf k}}}\sum_{nn'{\bf k}}|g^\nu_{n'{\bf k+q},n\bf k}|^2\nonumber\\ & \times & \delta(\epsilon_{n'{\bf k+\bf q}}-\epsilon_F)\delta(\epsilon_{n\bf k}-\epsilon_F),
\end{eqnarray} 
where $\epsilon_{n\bf k}$ is the energy of an electron in the band $n$ and wave number $\bf k$, $\omega_\nu({\bf q})$ the frequency of a phonon in the mode $\nu$ and wave number $\bf q$,  $\epsilon_F$ is the Fermi energy, $N_{{\bf k}}$ the number of ${\bf k}$ points in the sum, and $g^\nu_{n'{\bf k+q},n\bf k}$ are the electron-phonon matrix elements associated to the scattering of electrons with band energies $\epsilon_{n\bf k}$ and $\epsilon_{n'{\bf k}+{\bf q}}$ with a phonon of frequency $\omega_\nu({\bf q})$. Superconducting properties can be then calculated with the Eliashberg function
\begin{equation}
\alpha^2 F(\omega)=\frac{1}{2\pi N(\epsilon_F)N_{{\bf q}}}\sum_{\nu{\bf q}}\frac{\gamma_\nu({\bf q})}{\omega_\nu({\bf q})}\delta(\omega-\omega_\nu({\bf q})),
    \label{eq:a2f}
\end{equation}
where $N(\epsilon_F)$ is the density of states at the Fermi energy and $N_{{\bf q}}$ the number of $\bf q$ points in the sum. For instance, the electron-phonon coupling constant $\lambda$ is the $\omega\to\infty$ limit of the so-called integrated electron-phonon coupling constant
\begin{align}
\lambda(\omega)=2\int_0^\omega\frac{\alpha^2 F(\omega')}{\omega'}\mathrm d\omega'.
\end{align}
We have estimated $T_c$ by solving the Allen-Dynes equations \cite{eliashberg1960interactions,allen1983theory,allen1975transition}, choosing the effective Coulomb potential $\mu^*$ between $0.08$ and $0.12$.

The calculation of the electron-phonon properties has been done by making use in Eq. \eqref{linewidth} of a  60$\times$60$\times$1 grid of $\bf k$ points and an 8$\times$8$\times$1 grid of {\bf q} points in Eq. \eqref{eq:a2f}. To approximate the delta functions in Eq. \eqref{linewidth}, we have used Gaussians with a  broadening of $0.002$ Ry. The $\bf k$-point grid yielded converged $\lambda$ values even with such a small broadening of the Delta funcitons.

We have also studied the possibility of enhancing superconductivity by doping the system. We have doped the system by directly inserting or removing electrons and solving the Kohn-Sham equations, or including an external electric field in the calculations as it corresponds to a field effect transistor (FET) doping setup  \cite{brumme2014electrochemical,brumme2015first}. After doping, the crystal structure has been relaxed again until the forces and the strain tensor vanish. We have repeated the procedure subtracting electrons from the system. We have recomputed the electron and phonon bands, as well as the electron-phonon coupling with the inclusion of extra electrons or holes. The inclusion of the electric field in the calculations gave electronic band structures in agreement with those obtained with direct inclusion of electrons, validating the simpler approach. Therefore, here we report exclusively the results obtained by direct addition and subtraction of electrons.

\section{Results}

\label{results}

\subsection{Non doped case}

\subsubsection{Structure and bonding}

We describe the centered rectangular lattice of $Cmmm$ hydrogen boride in the primitive cell as shown in Fig. \ref{structure}, which includes two boron and two hydrogen atoms. In this description the two basis vectors $\bf a$ and $\bf b$ have the same length, which according to our DFT relaxations has a value of $a=b=5.739 a_0$. The angle between them is  $\gamma=59.33^\circ$. All boron atoms form a quasihexagonal honeycomb-like lattice and the hydrogen atoms sit at a distance $d$ from the boron plane. More precisely the boron atoms are at $\alpha(\boldsymbol a+\boldsymbol b)$ and $(1-\alpha)(\boldsymbol a+\boldsymbol b)$, while the hydrogen atoms sit at $1/2(\boldsymbol a+\boldsymbol b)\pm d\hat{\boldsymbol{k}}$, where $\hat{\boldsymbol{k}}$ is the unitary vector in the out-of-plane direction. According to our calculations $\alpha=0.328$ and $d=1.828a_0$. All the parameters that determine the structure are summarized in Table \ref{tabstruct}. The obtained structural parameters are in agreement with previous calculations \cite{kawamura2019photoinduced,tateishi2019semimetallicity}.

As illustrated by the calculated electron localization function (ELF) presented in Fig. \ref{elfnodop}, the presence of hydrogen affects the B-B bonding significantly. In a system like MgB$_2$, where the B layer is perfectly hexagonal, all boron atoms are covalently bonded at a distance of 3.34 $a_0$ \cite{tateishi2019semimetallicity}. The presence of hydrogen, however, changes the bonding pattern by shortening the B-B covalent that does not have a hydrogen atom above and below, while increasing the B-B length in the other bond due to the creation of a bridging B-H-B bond \cite{tominaka2020geometrical}. In fact, the presence of hydrogen destroys the covalent bond between the boron atoms by creating this bridging bonding. The B-B distances in the covalent bond are 3.25 $a_0$ while 3.44 $a_0$ between those affected by the B-H-B bonding.

\begin{table}
\begin{tabular}{c|c|c|c|c}
&$a$ $(a_0)$&$\alpha$&$\gamma$&$d$ $(a_0)$\\\hline
Non doped&5.739&0.328&59.33$^\circ$&1.828\\\hline
Doped electrons&5.739&0.326&59.36$^\circ$&1.810\\\hline
Doped holes&5.739&0.330&59.27$^\circ$&1.838
\end{tabular}
	\caption{Structural parameters of $Cmmm$ HB in the non doped and doped cases.}
	\label{tabstruct}
\end{table}

\subsubsection{Electronic bands and Fermi surface}

\label{elec}

The band structure and the density of states (DOS) projected onto atomic orbitals are plotted in Fig. \ref{bandsanddos}. The bands, which are in agreement with previous DFT and tight-binding results \cite{tateishi2019semimetallicity,kawamura2019photoinduced}, show a semimetallic character, with a hole pocket at $\Gamma$ and an electron pocket at Y, with two distinct bands crossing the Fermi level. The character of these two bands is mainly associated to the boron $p$ orbitals. The hole pocket at $\Gamma$ is coming from in-plane $p_x$ and $p_y$ orbitals, while the $Y$ pocket from boron $p_z$ \cite{tateishi2019semimetallicity}. 

The contribution of hydrogen to these bands that cross the Fermi level is scarce. Looking at the DOS projected onto atomic orbitals, we can see that  hydrogen contributes to the DOS at the Fermi level only a 1.6\%. The DOS at the Femi level is thus dominated by boron $p$ orbitals. Bands above approximately 2 eV above the Fermi level at $\Gamma$ do have, on the contrary, some considerable H character. 

\begin{figure*}
\centering
\includegraphics[width=\linewidth]{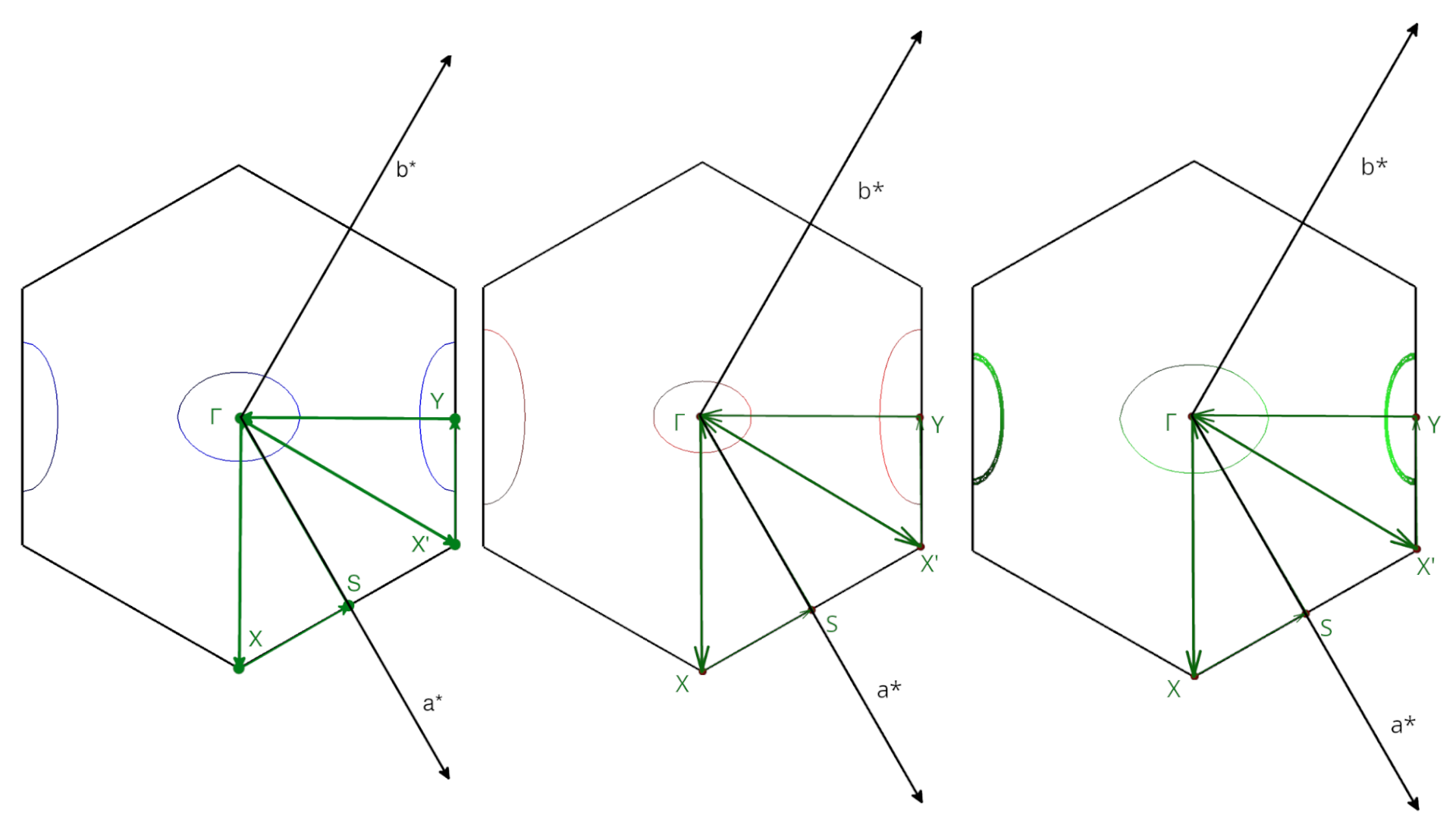}
\caption{Fermi surfaces for the non doped system (left), the one doped with electrons (center), and the one with holes (right) for $Cmmm$ HB. High-symmetry points are marked in green. Reciprocal lattice vectors ${\bf a}^*$ and ${\bf b}^*$ are included as well. The path used for the calculation of the electronic bands and phonon spectra is illustrated.}
\label{fermisurfs}
\end{figure*}

The hole pocket centered at $\Gamma$ and the electron pocket centered at Y are evident in the Fermi surface of Fig. \ref{fermisurfs}. Both pockets are ellipsoidal, but, due to the semimetallic character of HB, the area enclosed by the pockets is small. Indeed, the system is not far from an insulating state, as it has been suggested that a gap can be opened with a small strain \cite{mortazavi2018borophene}.

\subsubsection{Phonons and electron-phonon coupling}
\label{phon}

\begin{figure*}
\centering
\includegraphics[width=\textwidth]{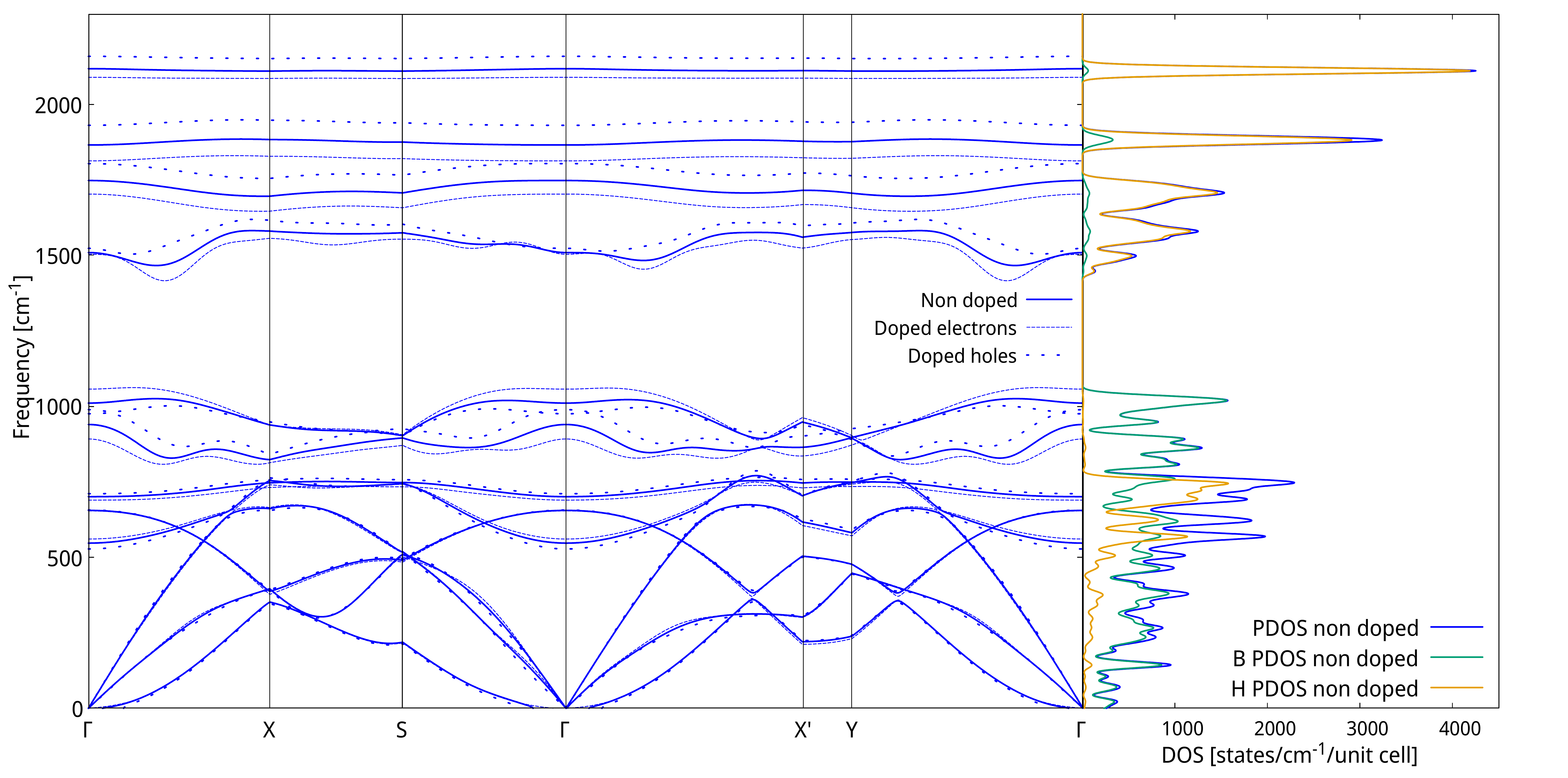}
\caption{Phonon spectrum (left panel) for the non doped $Cmmm$ HB, together with the doped cases with electrons and holes. 
The phonon density of states (PDOS) (right panel) is plotted, including also its decomposition onto different atoms, only for the non doped case. 
The PDOS decomposition is similar in the doped cases.
}
\label{phononbands}
\end{figure*}

We have calculated the phonon bands and the phonon DOS (PDOS), as well as the projection of the PDOS onto boron and hydrogen atoms (see Fig. \ref{phononbands}). As expected for a 2D material \cite{katsnelson2013graphene}, the flexural out-of-plane acoustic mode has a quadratic dispersion due to rotational symmetry. The phonon spectrum has three distinctive regions. Below  $\sim$700 cm$^{-1}$ there are three acoustic and another three optical modes. These modes show, in general, a mixed character between hydrogen and boron. The hydrogen character of these low-energy modes comes from the vibrations of H atoms along the $y$ direction. At higher energies, between approximately 750 cm$^{-1}$ and 1100 cm$^{-1}$, there are two isolated phonon bands that describe in-plane boron vibrations, reminiscent of the $E_{2g}$ modes that are strongly coupled to the electrons in the similar MgB$_2$ \cite{choi2002origin,an2001superconductivity,yildirim2001giant}. For comparison, the energy of the $E_{2g}$ mode in MgB$_2$ is around 600 cm$^{-1}$ \cite{astuto2007weak}, smaller than the analogous modes in HB. The remaining four phonon modes appear after a large energy gap above 1500 cm$^{-1}$ and have a dominant hydrogen character, with vibrations along the $x$ and $z$ directions.

As we can observe in Fig. \ref{alpha2f}, where the Eliashberg function $\alpha^2F(\omega)$ and the integrated electron-phonon coupling constant $\lambda(\omega)$ are shown, phonon modes with energies between $500$ and $1000$ cm$^{-1}$ contribute more to the electron-phonon coupling. The modes mainly contributing to the peaks in this energy range  involve boron and hydrogen displacements, as well as the in-plane pure boron modes, analogous to the $E_{2g}$ of MgB$_2$. There is another significant peak at around $1500$ cm$^{-1}$, related to hydrogen-character modes, but, as expected, the lower frequency modes are the ones that contribute most to the electron-phonon coupling.  Unfortunately, the integrated electron-phonon coupling $\lambda$ reaches a very low value of approximately $0.2$, which yields a very low superconducting $T_c$ of only 11 mK.

\subsection{Doped cases}

The DOS of 2D materials is energy independent in the free electron-like limit and, thus, since $\lambda$ is proportional to $N(\epsilon_F)$, doping is expected to not affect the electron-phonon coupling constant. However, there are examples in the literature where electrostatic doping has been used to induce superconductivity in two-dimensional materials such as amorphous bismuth \cite{parendo2006electrostatic} or the LaIO$_3$/SrTiO$_3$ interface \cite{caviglia2008electric}, in analogy to the common behavior of the cuprates \cite{ahn1999electrostatic}. Also, this FET doping mechanism has been used to induce and tune supercconductivity in transition metal dichalcogenides \cite{saito2016highly}, for instance in MoS$_2$, where doping induces first an insulator to metal transition, and a superconducting state at further electron doping, with a $T_c$ that can reach a value as high as 10 K \cite{saito2015metallic}. 

Motivated by these experimental results and the possibility of FET doping two-dimensional materials in the laboratory, we have studied the effect of both electron and hole doping in the structural, electronic, phononic, and superconducting properties of $Cmmm$ HB, with the hope that its low $T_c$ may be enhanced. In the following we report the results obtained with electron and hole doping, which will be referred as the electron/holes doped cases respectively. We dope the system with $0.075$ electrons(holes) per unit cell, which corresponds to a $9.39\times10^{13}$ electrons(holes)/cm$^2$.

\subsubsection{Structure and bonding}

By relaxing the structure before with DFT in the doped system, we can determine how the structure and the bonding are affected when extra electrons or holes are included in the system. The modified structural parameters are summarized in Table \ref{tabstruct}. Doping leaves the lattice parameter $a$ unvaried, but increases the $\gamma$ angle when electrons are added and decreases it if, instead, extra holes are present. Similarly, the distance of hydrogen from the plane decreases in the electron doped case, while increases in the hole doped case. These results suggest that electron doping strengthens the B-B and B-H-B covalent bonds.

\subsubsection{Electronic bands and Fermi surface}

We report the electronic bands, together with the DOS, in Fig. \ref{bandsanddos}. We observe that, as expected, the conduction bands of the electron doped case tend to go down with respect to the non doped case, while the bands of the holes doping tend to go up with respect to the non doped case. The effect is that in the electron doped case the hole pocket at $\Gamma$ decreases and the electron pocket at Y increases. The opposite happens in the hole doped case. This effect is clear in the Fermi surface in Fig. \ref{fermisurfs}. The character of the bands is barely affected by the doping and the percentage of H states in the DOS at the Fermi energy remains minimal, as in the non doped case. Indeed, with the doping values assumed, $N(\epsilon_F)$ remains practically unchanged, which is reasonable given the constant DOS in the vicinity of the Fermi level.

\subsubsection{Phonons and electron-phonon coupling}

As we can see in Fig. \ref{phononbands}, electron or hole doping barely affects the lowest six phonon modes. The differences become more appreciable for the isolated two boron in-plane bands in the frequency range between 750 and 1100 cm$^{-1}$. Electron doping increases the splitting between these two modes, while whole doping decreases it. Finally, the hydrogen-dominated modes at high energies are softened in the electron doped case and hardened in the hole doped case.

\begin{figure}
\centering
\includegraphics[width=\columnwidth]{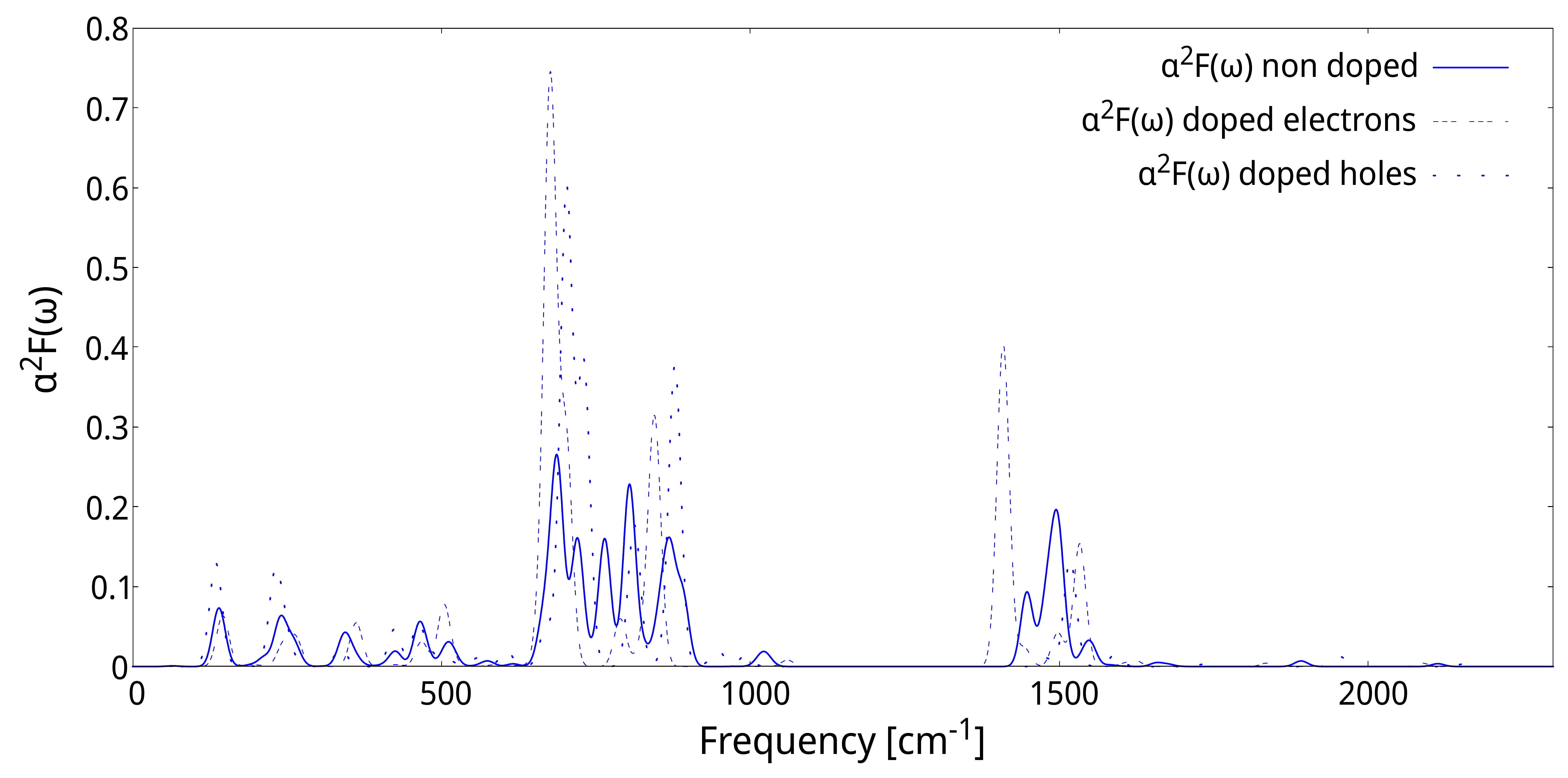}
\includegraphics[width=\columnwidth]{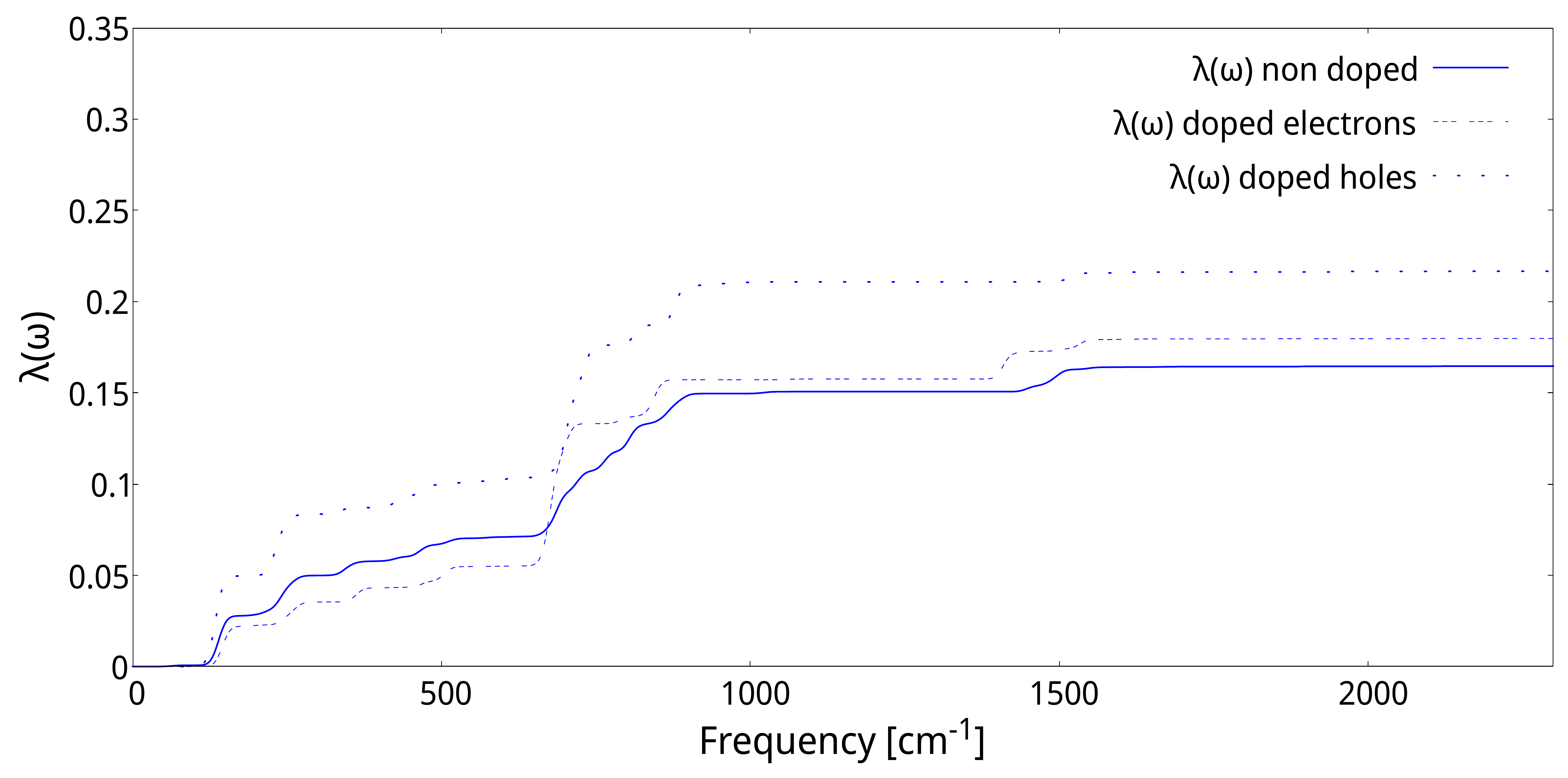}
\caption{Eliashberg function $\alpha^2F(\omega)$ and integrated electron-phonon coupling constant $\lambda(\omega)$ for the non doped system, as well as the doped systems with electrons and holes.}
\label{alpha2f}
\end{figure}

Taking a look at the values of the Eliashberg function $\alpha^2F(\omega)$ and the integrated electron-phonon coupling constant $\lambda(\omega)$, plotted in Fig. \ref{alpha2f},  we can see that in the doped cases still the modes in the 500-1000 cm$^{-1}$ range are those with the most important contribution to $\lambda$. However, specially the low-energy acoustic modes, are more coupled to the electrons in the whole doped case. Also the modes in the 500-1000 cm$^{-1}$ range seem to be more prone to the electrons in the doped cases. We attribute these differences to the change in the electron-phonon coupling matrix elements and not to the shift of the phonon frequencies. The slight changes in the bonding pattern can explain a subtle change of the electron-phonon matrix elements themselves. Nevertheless, doping, either hole or electron, is not able to enhance $\lambda$ sufficiently to reach a sizable critical temperature in HB. The value of $T_c$ in the hole doped case is 21 mK. Thus, despite the opposite is the case in other systems, doping does not seem to be an efficient strategy to increase the critical temperatures in hydrogen boride, at least, in the $Cmmm$ phase synthesized experimentally.

\section{Discussion}

\label{discussion}

Neither the non doped nor the doped HB present a sizable superconductivity in the $Cmmm$ phase. This can be traced to a variety of reasons. It has been recently argue \cite{belli2021strong} that hydrogen-based compounds can reach a high $T_c$ when an electronic bonding network between localized units is created. The \emph{networking value} $\phi$, which was defined as the highest value of the ELF that creates an isosurface spanning through the whole crystal in all three Cartesian directions, measures precisely that network. Interestingly, $\phi$ correlates better than any other descriptor with $T_c$. By multiplying the networking value with the fraction of hydrogen in the compound $H_f$ (0.5 for HB) and with the third root of the hydrogen fraction of the total DOS at the Fermi energy $H_{\mathrm{DOS}}$, i.e. defining $\Phi_{\mathrm{DOS}}=\phi H_f \sqrt[3]{H_{\mathrm{DOS}}}$,  a rather good correlation was obtained between $\Phi_{\mathrm{DOS}}$ and $T_c$, capable of estimating the critical temperature with the empirical \cite{belli2021strong}
\begin{equation}
    T_c = (750\Phi_{\mathrm{DOS}}-85) \mathrm{K}
    \label{eq:networking}
\end{equation} equation . Here we assume that the networking value in a 2D system should be determined by the highest value of the ELF that creates an isosurface that expands in the whole crystal in the plane. With this adapted definition we obtain  $\phi=0.53$, a rather large value. However, the fact that the hydrogen fraction of the DOS at the Fermi level is so low in this system suppresses all the $T_c$ according to Eq. \eqref{eq:networking}, both in the undoped and doped cases.

Furthermore, the presence of hydrogen alters the bonding of the boron layer creating the B-H-B bridges. This fact suppresses the electron-phonon coupling of the boron in-plane modes due to the large impact that the presence of hydrogen has in the band structure \cite{tateishi2019semimetallicity,kawamura2019photoinduced}. Therefore, the presence of hydrogen also suppresses the electron-phonon coupling in the boron layer itself. Hydrogen doping of borophene is not thus favoring the emergence of high-temperature superconductivity in hydrogen boride, it has the opposite effect at least in the $Cmmm$ phase.

\section{Conclusions}

\label{conclusion}

From the results of our analysis we can conclude that, despite the large expectations in the literature \cite{tateishi2019semimetallicity}, hydrogen boride is not a strongly coupled superconductor mainly because of the weak hydrogen character of the density of states at the Fermi level and the effect of the presence of hydrogen in the bonding of the boron layer. The $T_c$ values are not sizable, below 1 K, even if the system is doped. Our results suggest that hydrogenated monolayers will be high-temperature superconductors, comparable in their $T_c$'s to high-pressure superhydrides, only if the electronic states at the Fermi level have a large hydrogen character.

\section*{Acknowledgements}

We have received funding from the European Research Council (ERC) under the European Union’s Horizon 2020 research and innovation programme (grant agreement No 802533).

\bibliographystyle{apsrev4-2.bst}
\bibliography{biblio}

\begin{thebibliography}{42}%
\makeatletter
\providecommand \@ifxundefined [1]{%
 \@ifx{#1\undefined}
}%
\providecommand \@ifnum [1]{%
 \ifnum #1\expandafter \@firstoftwo
 \else \expandafter \@secondoftwo
 \fi
}%
\providecommand \@ifx [1]{%
 \ifx #1\expandafter \@firstoftwo
 \else \expandafter \@secondoftwo
 \fi
}%
\providecommand \natexlab [1]{#1}%
\providecommand \enquote  [1]{``#1''}%
\providecommand \bibnamefont  [1]{#1}%
\providecommand \bibfnamefont [1]{#1}%
\providecommand \citenamefont [1]{#1}%
\providecommand \href@noop [0]{\@secondoftwo}%
\providecommand \href [0]{\begingroup \@sanitize@url \@href}%
\providecommand \@href[1]{\@@startlink{#1}\@@href}%
\providecommand \@@href[1]{\endgroup#1\@@endlink}%
\providecommand \@sanitize@url [0]{\catcode `\\12\catcode `\$12\catcode
  `\&12\catcode `\#12\catcode `\^12\catcode `\_12\catcode `\%12\relax}%
\providecommand \@@startlink[1]{}%
\providecommand \@@endlink[0]{}%
\providecommand \url  [0]{\begingroup\@sanitize@url \@url }%
\providecommand \@url [1]{\endgroup\@href {#1}{\urlprefix }}%
\providecommand \urlprefix  [0]{URL }%
\providecommand \Eprint [0]{\href }%
\providecommand \doibase [0]{https://doi.org/}%
\providecommand \selectlanguage [0]{\@gobble}%
\providecommand \bibinfo  [0]{\@secondoftwo}%
\providecommand \bibfield  [0]{\@secondoftwo}%
\providecommand \translation [1]{[#1]}%
\providecommand \BibitemOpen [0]{}%
\providecommand \bibitemStop [0]{}%
\providecommand \bibitemNoStop [0]{.\EOS\space}%
\providecommand \EOS [0]{\spacefactor3000\relax}%
\providecommand \BibitemShut  [1]{\csname bibitem#1\endcsname}%
\let\auto@bib@innerbib\@empty
\bibitem [{\citenamefont {Drozdov}\ \emph {et~al.}(2015)\citenamefont
  {Drozdov}, \citenamefont {Eremets}, \citenamefont {Troyan}, \citenamefont
  {Ksenofontov},\ and\ \citenamefont {Shylin}}]{drozdov2015conventional}%
  \BibitemOpen
  \bibfield  {author} {\bibinfo {author} {\bibfnamefont {A.}~\bibnamefont
  {Drozdov}}, \bibinfo {author} {\bibfnamefont {M.}~\bibnamefont {Eremets}},
  \bibinfo {author} {\bibfnamefont {I.}~\bibnamefont {Troyan}}, \bibinfo
  {author} {\bibfnamefont {V.}~\bibnamefont {Ksenofontov}},\ and\ \bibinfo
  {author} {\bibfnamefont {S.~I.}\ \bibnamefont {Shylin}},\ }\href@noop {}
  {\bibfield  {journal} {\bibinfo  {journal} {Nature}\ }\textbf {\bibinfo
  {volume} {525}},\ \bibinfo {pages} {73} (\bibinfo {year} {2015})}\BibitemShut
  {NoStop}%
\bibitem [{\citenamefont {Somayazulu}\ \emph {et~al.}(2019)\citenamefont
  {Somayazulu}, \citenamefont {Ahart}, \citenamefont {Mishra}, \citenamefont
  {Geballe}, \citenamefont {Baldini}, \citenamefont {Meng}, \citenamefont
  {Struzhkin},\ and\ \citenamefont {Hemley}}]{somayazulu2019evidence}%
  \BibitemOpen
  \bibfield  {author} {\bibinfo {author} {\bibfnamefont {M.}~\bibnamefont
  {Somayazulu}}, \bibinfo {author} {\bibfnamefont {M.}~\bibnamefont {Ahart}},
  \bibinfo {author} {\bibfnamefont {A.~K.}\ \bibnamefont {Mishra}}, \bibinfo
  {author} {\bibfnamefont {Z.~M.}\ \bibnamefont {Geballe}}, \bibinfo {author}
  {\bibfnamefont {M.}~\bibnamefont {Baldini}}, \bibinfo {author} {\bibfnamefont
  {Y.}~\bibnamefont {Meng}}, \bibinfo {author} {\bibfnamefont {V.~V.}\
  \bibnamefont {Struzhkin}},\ and\ \bibinfo {author} {\bibfnamefont {R.~J.}\
  \bibnamefont {Hemley}},\ }\href@noop {} {\bibfield  {journal} {\bibinfo
  {journal} {Physical Review Letters}\ }\textbf {\bibinfo {volume} {122}},\
  \bibinfo {pages} {027001} (\bibinfo {year} {2019})}\BibitemShut {NoStop}%
\bibitem [{\citenamefont {Drozdov}\ \emph {et~al.}(2019)\citenamefont
  {Drozdov}, \citenamefont {Kong}, \citenamefont {Minkov}, \citenamefont
  {Besedin}, \citenamefont {Kuzovnikov}, \citenamefont {Mozaffari},
  \citenamefont {Balicas}, \citenamefont {Balakirev}, \citenamefont {Graf},
  \citenamefont {Prakapenka} \emph {et~al.}}]{drozdov2019superconductivity}%
  \BibitemOpen
  \bibfield  {author} {\bibinfo {author} {\bibfnamefont {A.}~\bibnamefont
  {Drozdov}}, \bibinfo {author} {\bibfnamefont {P.}~\bibnamefont {Kong}},
  \bibinfo {author} {\bibfnamefont {V.}~\bibnamefont {Minkov}}, \bibinfo
  {author} {\bibfnamefont {S.}~\bibnamefont {Besedin}}, \bibinfo {author}
  {\bibfnamefont {M.}~\bibnamefont {Kuzovnikov}}, \bibinfo {author}
  {\bibfnamefont {S.}~\bibnamefont {Mozaffari}}, \bibinfo {author}
  {\bibfnamefont {L.}~\bibnamefont {Balicas}}, \bibinfo {author} {\bibfnamefont
  {F.}~\bibnamefont {Balakirev}}, \bibinfo {author} {\bibfnamefont
  {D.}~\bibnamefont {Graf}}, \bibinfo {author} {\bibfnamefont {V.}~\bibnamefont
  {Prakapenka}}, \emph {et~al.},\ }\href@noop {} {\bibfield  {journal}
  {\bibinfo  {journal} {Nature}\ }\textbf {\bibinfo {volume} {569}},\ \bibinfo
  {pages} {528} (\bibinfo {year} {2019})}\BibitemShut {NoStop}%
\bibitem [{\citenamefont {Snider}\ \emph {et~al.}(2021)\citenamefont {Snider},
  \citenamefont {Dasenbrock-Gammon}, \citenamefont {McBride}, \citenamefont
  {Wang}, \citenamefont {Meyers}, \citenamefont {Lawler}, \citenamefont
  {Zurek}, \citenamefont {Salamat},\ and\ \citenamefont
  {Dias}}]{snider2021synthesis}%
  \BibitemOpen
  \bibfield  {author} {\bibinfo {author} {\bibfnamefont {E.}~\bibnamefont
  {Snider}}, \bibinfo {author} {\bibfnamefont {N.}~\bibnamefont
  {Dasenbrock-Gammon}}, \bibinfo {author} {\bibfnamefont {R.}~\bibnamefont
  {McBride}}, \bibinfo {author} {\bibfnamefont {X.}~\bibnamefont {Wang}},
  \bibinfo {author} {\bibfnamefont {N.}~\bibnamefont {Meyers}}, \bibinfo
  {author} {\bibfnamefont {K.~V.}\ \bibnamefont {Lawler}}, \bibinfo {author}
  {\bibfnamefont {E.}~\bibnamefont {Zurek}}, \bibinfo {author} {\bibfnamefont
  {A.}~\bibnamefont {Salamat}},\ and\ \bibinfo {author} {\bibfnamefont {R.~P.}\
  \bibnamefont {Dias}},\ }\href
  {https://doi.org/10.1103/PhysRevLett.126.117003} {\bibfield  {journal}
  {\bibinfo  {journal} {Phys. Rev. Lett.}\ }\textbf {\bibinfo {volume} {126}},\
  \bibinfo {pages} {117003} (\bibinfo {year} {2021})}\BibitemShut {NoStop}%
\bibitem [{\citenamefont {Kong}\ \emph {et~al.}(2021)\citenamefont {Kong},
  \citenamefont {Minkov}, \citenamefont {Kuzovnikov}, \citenamefont {Drozdov},
  \citenamefont {Besedin}, \citenamefont {Mozaffari}, \citenamefont {Balicas},
  \citenamefont {Balakirev}, \citenamefont {Prakapenka}, \citenamefont
  {Chariton}, \citenamefont {Knyazev}, \citenamefont {Greenberg},\ and\
  \citenamefont {Eremets}}]{kong2021superconductivity}%
  \BibitemOpen
  \bibfield  {author} {\bibinfo {author} {\bibfnamefont {P.}~\bibnamefont
  {Kong}}, \bibinfo {author} {\bibfnamefont {V.~S.}\ \bibnamefont {Minkov}},
  \bibinfo {author} {\bibfnamefont {M.~A.}\ \bibnamefont {Kuzovnikov}},
  \bibinfo {author} {\bibfnamefont {A.~P.}\ \bibnamefont {Drozdov}}, \bibinfo
  {author} {\bibfnamefont {S.~P.}\ \bibnamefont {Besedin}}, \bibinfo {author}
  {\bibfnamefont {S.}~\bibnamefont {Mozaffari}}, \bibinfo {author}
  {\bibfnamefont {L.}~\bibnamefont {Balicas}}, \bibinfo {author} {\bibfnamefont
  {F.~F.}\ \bibnamefont {Balakirev}}, \bibinfo {author} {\bibfnamefont {V.~B.}\
  \bibnamefont {Prakapenka}}, \bibinfo {author} {\bibfnamefont
  {S.}~\bibnamefont {Chariton}}, \bibinfo {author} {\bibfnamefont {D.~A.}\
  \bibnamefont {Knyazev}}, \bibinfo {author} {\bibfnamefont {E.}~\bibnamefont
  {Greenberg}},\ and\ \bibinfo {author} {\bibfnamefont {M.~I.}\ \bibnamefont
  {Eremets}},\ }\href {https://doi.org/10.1038/s41467-021-25372-2} {\bibfield
  {journal} {\bibinfo  {journal} {Nature Communications}\ }\textbf {\bibinfo
  {volume} {12}},\ \bibinfo {pages} {5075} (\bibinfo {year}
  {2021})}\BibitemShut {NoStop}%
\bibitem [{\citenamefont {Semenok}\ \emph {et~al.}(2020)\citenamefont
  {Semenok}, \citenamefont {Troyan}, \citenamefont {Kvashnin}, \citenamefont
  {Ivanova}, \citenamefont {Hanfland}, \citenamefont {Sadakov}, \citenamefont
  {Sobolevskiy}, \citenamefont {Pervakov}, \citenamefont {Gavriliuk},
  \citenamefont {Lyubutin} \emph {et~al.}}]{semenok2020superconductivity}%
  \BibitemOpen
  \bibfield  {author} {\bibinfo {author} {\bibfnamefont {D.~V.}\ \bibnamefont
  {Semenok}}, \bibinfo {author} {\bibfnamefont {I.~A.}\ \bibnamefont {Troyan}},
  \bibinfo {author} {\bibfnamefont {A.~G.}\ \bibnamefont {Kvashnin}}, \bibinfo
  {author} {\bibfnamefont {A.~G.}\ \bibnamefont {Ivanova}}, \bibinfo {author}
  {\bibfnamefont {M.}~\bibnamefont {Hanfland}}, \bibinfo {author}
  {\bibfnamefont {A.~V.}\ \bibnamefont {Sadakov}}, \bibinfo {author}
  {\bibfnamefont {O.~A.}\ \bibnamefont {Sobolevskiy}}, \bibinfo {author}
  {\bibfnamefont {K.~S.}\ \bibnamefont {Pervakov}}, \bibinfo {author}
  {\bibfnamefont {A.~G.}\ \bibnamefont {Gavriliuk}}, \bibinfo {author}
  {\bibfnamefont {I.~S.}\ \bibnamefont {Lyubutin}}, \emph {et~al.},\
  }\href@noop {} {\bibfield  {journal} {\bibinfo  {journal} {arXiv preprint
  arXiv:2012.04787}\ } (\bibinfo {year} {2020})}\BibitemShut {NoStop}%
\bibitem [{\citenamefont
  {Skoskiewicz}(1972)}]{skoskiewicz1972superconductivity}%
  \BibitemOpen
  \bibfield  {author} {\bibinfo {author} {\bibfnamefont {T.}~\bibnamefont
  {Skoskiewicz}},\ }\href@noop {} {\bibfield  {journal} {\bibinfo  {journal}
  {physica status solidi (a)}\ }\textbf {\bibinfo {volume} {11}},\ \bibinfo
  {pages} {K123} (\bibinfo {year} {1972})}\BibitemShut {NoStop}%
\bibitem [{\citenamefont {Satterthwaite}\ and\ \citenamefont
  {Toepke}(1970)}]{satterthwaite1970superconductivity}%
  \BibitemOpen
  \bibfield  {author} {\bibinfo {author} {\bibfnamefont {C.}~\bibnamefont
  {Satterthwaite}}\ and\ \bibinfo {author} {\bibfnamefont {I.}~\bibnamefont
  {Toepke}},\ }\href@noop {} {\bibfield  {journal} {\bibinfo  {journal}
  {Physical Review Letters}\ }\textbf {\bibinfo {volume} {25}},\ \bibinfo
  {pages} {741} (\bibinfo {year} {1970})}\BibitemShut {NoStop}%
\bibitem [{\citenamefont {Nagamatsu}\ \emph {et~al.}(2001)\citenamefont
  {Nagamatsu}, \citenamefont {Nakagawa}, \citenamefont {Muranaka},
  \citenamefont {Zenitani},\ and\ \citenamefont
  {Akimitsu}}]{nagamatsu_superconductivity_2001}%
  \BibitemOpen
  \bibfield  {author} {\bibinfo {author} {\bibfnamefont {J.}~\bibnamefont
  {Nagamatsu}}, \bibinfo {author} {\bibfnamefont {N.}~\bibnamefont {Nakagawa}},
  \bibinfo {author} {\bibfnamefont {T.}~\bibnamefont {Muranaka}}, \bibinfo
  {author} {\bibfnamefont {Y.}~\bibnamefont {Zenitani}},\ and\ \bibinfo
  {author} {\bibfnamefont {J.}~\bibnamefont {Akimitsu}},\ }\href
  {https://doi.org/10.1038/35065039} {\bibfield  {journal} {\bibinfo  {journal}
  {Nature}\ }\textbf {\bibinfo {volume} {410}},\ \bibinfo {pages} {63}
  (\bibinfo {year} {2001})}\BibitemShut {NoStop}%
\bibitem [{\citenamefont {Bekaert}\ \emph {et~al.}(2019)\citenamefont
  {Bekaert}, \citenamefont {Petrov}, \citenamefont {Aperis}, \citenamefont
  {Oppeneer},\ and\ \citenamefont {Milo{\v{s}}evi{\'c}}}]{bekaert2019hydrogen}%
  \BibitemOpen
  \bibfield  {author} {\bibinfo {author} {\bibfnamefont {J.}~\bibnamefont
  {Bekaert}}, \bibinfo {author} {\bibfnamefont {M.}~\bibnamefont {Petrov}},
  \bibinfo {author} {\bibfnamefont {A.}~\bibnamefont {Aperis}}, \bibinfo
  {author} {\bibfnamefont {P.~M.}\ \bibnamefont {Oppeneer}},\ and\ \bibinfo
  {author} {\bibfnamefont {M.}~\bibnamefont {Milo{\v{s}}evi{\'c}}},\
  }\href@noop {} {\bibfield  {journal} {\bibinfo  {journal} {Physical review
  letters}\ }\textbf {\bibinfo {volume} {123}},\ \bibinfo {pages} {077001}
  (\bibinfo {year} {2019})}\BibitemShut {NoStop}%
\bibitem [{\citenamefont {Savini}\ \emph {et~al.}(2010)\citenamefont {Savini},
  \citenamefont {Ferrari},\ and\ \citenamefont {Giustino}}]{savini2010first}%
  \BibitemOpen
  \bibfield  {author} {\bibinfo {author} {\bibfnamefont {G.}~\bibnamefont
  {Savini}}, \bibinfo {author} {\bibfnamefont {A.}~\bibnamefont {Ferrari}},\
  and\ \bibinfo {author} {\bibfnamefont {F.}~\bibnamefont {Giustino}},\
  }\href@noop {} {\bibfield  {journal} {\bibinfo  {journal} {Physical review
  letters}\ }\textbf {\bibinfo {volume} {105}},\ \bibinfo {pages} {037002}
  (\bibinfo {year} {2010})}\BibitemShut {NoStop}%
\bibitem [{\citenamefont {Nishino}\ \emph {et~al.}(2017)\citenamefont
  {Nishino}, \citenamefont {Fujita}, \citenamefont {Cuong}, \citenamefont
  {Tominaka}, \citenamefont {Miyauchi}, \citenamefont {Iimura}, \citenamefont
  {Hirata}, \citenamefont {Umezawa}, \citenamefont {Okada}, \citenamefont
  {Nishibori}, \citenamefont {Fujino}, \citenamefont {Fujimori}, \citenamefont
  {Ito}, \citenamefont {Nakamura}, \citenamefont {Hosono},\ and\ \citenamefont
  {Kondo}}]{nishino2017formation}%
  \BibitemOpen
  \bibfield  {author} {\bibinfo {author} {\bibfnamefont {H.}~\bibnamefont
  {Nishino}}, \bibinfo {author} {\bibfnamefont {T.}~\bibnamefont {Fujita}},
  \bibinfo {author} {\bibfnamefont {N.~T.}\ \bibnamefont {Cuong}}, \bibinfo
  {author} {\bibfnamefont {S.}~\bibnamefont {Tominaka}}, \bibinfo {author}
  {\bibfnamefont {M.}~\bibnamefont {Miyauchi}}, \bibinfo {author}
  {\bibfnamefont {S.}~\bibnamefont {Iimura}}, \bibinfo {author} {\bibfnamefont
  {A.}~\bibnamefont {Hirata}}, \bibinfo {author} {\bibfnamefont
  {N.}~\bibnamefont {Umezawa}}, \bibinfo {author} {\bibfnamefont
  {S.}~\bibnamefont {Okada}}, \bibinfo {author} {\bibfnamefont
  {E.}~\bibnamefont {Nishibori}}, \bibinfo {author} {\bibfnamefont
  {A.}~\bibnamefont {Fujino}}, \bibinfo {author} {\bibfnamefont
  {T.}~\bibnamefont {Fujimori}}, \bibinfo {author} {\bibfnamefont {S.-i.}\
  \bibnamefont {Ito}}, \bibinfo {author} {\bibfnamefont {J.}~\bibnamefont
  {Nakamura}}, \bibinfo {author} {\bibfnamefont {H.}~\bibnamefont {Hosono}},\
  and\ \bibinfo {author} {\bibfnamefont {T.}~\bibnamefont {Kondo}},\ }\href
  {https://doi.org/10.1021/jacs.7b06153} {\bibfield  {journal} {\bibinfo
  {journal} {Journal of the American Chemical Society}\ }\textbf {\bibinfo
  {volume} {139}},\ \bibinfo {pages} {13761} (\bibinfo {year} {2017})},\
  \bibinfo {note} {pMID: 28926230},\ \Eprint
  {https://arxiv.org/abs/https://doi.org/10.1021/jacs.7b06153}
  {https://doi.org/10.1021/jacs.7b06153} \BibitemShut {NoStop}%
\bibitem [{\citenamefont {Tateishi}\ \emph {et~al.}(2019)\citenamefont
  {Tateishi}, \citenamefont {Cuong}, \citenamefont {Moura}, \citenamefont
  {Cameau}, \citenamefont {Ishibiki}, \citenamefont {Fujino}, \citenamefont
  {Okada}, \citenamefont {Yamamoto}, \citenamefont {Araki}, \citenamefont {Ito}
  \emph {et~al.}}]{tateishi2019semimetallicity}%
  \BibitemOpen
  \bibfield  {author} {\bibinfo {author} {\bibfnamefont {I.}~\bibnamefont
  {Tateishi}}, \bibinfo {author} {\bibfnamefont {N.}~\bibnamefont {Cuong}},
  \bibinfo {author} {\bibfnamefont {C.}~\bibnamefont {Moura}}, \bibinfo
  {author} {\bibfnamefont {M.}~\bibnamefont {Cameau}}, \bibinfo {author}
  {\bibfnamefont {R.}~\bibnamefont {Ishibiki}}, \bibinfo {author}
  {\bibfnamefont {A.}~\bibnamefont {Fujino}}, \bibinfo {author} {\bibfnamefont
  {S.}~\bibnamefont {Okada}}, \bibinfo {author} {\bibfnamefont
  {A.}~\bibnamefont {Yamamoto}}, \bibinfo {author} {\bibfnamefont
  {M.}~\bibnamefont {Araki}}, \bibinfo {author} {\bibfnamefont
  {S.}~\bibnamefont {Ito}}, \emph {et~al.},\ }\href@noop {} {\bibfield
  {journal} {\bibinfo  {journal} {Physical Review Materials}\ }\textbf
  {\bibinfo {volume} {3}},\ \bibinfo {pages} {024004} (\bibinfo {year}
  {2019})}\BibitemShut {NoStop}%
\bibitem [{\citenamefont {Kawamura}\ \emph {et~al.}(2019)\citenamefont
  {Kawamura}, \citenamefont {Cuong}, \citenamefont {Fujita}, \citenamefont
  {Ishibiki}, \citenamefont {Hirabayashi}, \citenamefont {Yamaguchi},
  \citenamefont {Matsuda}, \citenamefont {Okada}, \citenamefont {Kondo},\ and\
  \citenamefont {Miyauchi}}]{kawamura2019photoinduced}%
  \BibitemOpen
  \bibfield  {author} {\bibinfo {author} {\bibfnamefont {R.}~\bibnamefont
  {Kawamura}}, \bibinfo {author} {\bibfnamefont {N.~T.}\ \bibnamefont {Cuong}},
  \bibinfo {author} {\bibfnamefont {T.}~\bibnamefont {Fujita}}, \bibinfo
  {author} {\bibfnamefont {R.}~\bibnamefont {Ishibiki}}, \bibinfo {author}
  {\bibfnamefont {T.}~\bibnamefont {Hirabayashi}}, \bibinfo {author}
  {\bibfnamefont {A.}~\bibnamefont {Yamaguchi}}, \bibinfo {author}
  {\bibfnamefont {I.}~\bibnamefont {Matsuda}}, \bibinfo {author} {\bibfnamefont
  {S.}~\bibnamefont {Okada}}, \bibinfo {author} {\bibfnamefont
  {T.}~\bibnamefont {Kondo}},\ and\ \bibinfo {author} {\bibfnamefont
  {M.}~\bibnamefont {Miyauchi}},\ }\href@noop {} {\bibfield  {journal}
  {\bibinfo  {journal} {Nature communications}\ }\textbf {\bibinfo {volume}
  {10}},\ \bibinfo {pages} {1} (\bibinfo {year} {2019})}\BibitemShut {NoStop}%
\bibitem [{\citenamefont {Jiao}\ \emph {et~al.}(2016)\citenamefont {Jiao},
  \citenamefont {Ma}, \citenamefont {Bell}, \citenamefont {Bilic},\ and\
  \citenamefont {Du}}]{jiao2016two-dimensional}%
  \BibitemOpen
  \bibfield  {author} {\bibinfo {author} {\bibfnamefont {Y.}~\bibnamefont
  {Jiao}}, \bibinfo {author} {\bibfnamefont {F.}~\bibnamefont {Ma}}, \bibinfo
  {author} {\bibfnamefont {J.}~\bibnamefont {Bell}}, \bibinfo {author}
  {\bibfnamefont {A.}~\bibnamefont {Bilic}},\ and\ \bibinfo {author}
  {\bibfnamefont {A.}~\bibnamefont {Du}},\ }\href
  {https://doi.org/https://doi.org/10.1002/anie.201604369} {\bibfield
  {journal} {\bibinfo  {journal} {Angewandte Chemie International Edition}\
  }\textbf {\bibinfo {volume} {55}},\ \bibinfo {pages} {10292} (\bibinfo {year}
  {2016})},\ \Eprint
  {https://arxiv.org/abs/https://onlinelibrary.wiley.com/doi/pdf/10.1002/anie.201604369}
  {https://onlinelibrary.wiley.com/doi/pdf/10.1002/anie.201604369} \BibitemShut
  {NoStop}%
\bibitem [{\citenamefont {Li}\ \emph {et~al.}(2021)\citenamefont {Li},
  \citenamefont {Kolluru}, \citenamefont {Rahn}, \citenamefont {Schwenker},
  \citenamefont {Li}, \citenamefont {Hennig}, \citenamefont {Darancet},
  \citenamefont {Chan},\ and\ \citenamefont {Hersam}}]{li2021synthesis}%
  \BibitemOpen
  \bibfield  {author} {\bibinfo {author} {\bibfnamefont {Q.}~\bibnamefont
  {Li}}, \bibinfo {author} {\bibfnamefont {V.~S.~C.}\ \bibnamefont {Kolluru}},
  \bibinfo {author} {\bibfnamefont {M.~S.}\ \bibnamefont {Rahn}}, \bibinfo
  {author} {\bibfnamefont {E.}~\bibnamefont {Schwenker}}, \bibinfo {author}
  {\bibfnamefont {S.}~\bibnamefont {Li}}, \bibinfo {author} {\bibfnamefont
  {R.~G.}\ \bibnamefont {Hennig}}, \bibinfo {author} {\bibfnamefont
  {P.}~\bibnamefont {Darancet}}, \bibinfo {author} {\bibfnamefont {M.~K.~Y.}\
  \bibnamefont {Chan}},\ and\ \bibinfo {author} {\bibfnamefont {M.~C.}\
  \bibnamefont {Hersam}},\ }\href {https://doi.org/10.1126/science.abg1874}
  {\bibfield  {journal} {\bibinfo  {journal} {Science}\ }\textbf {\bibinfo
  {volume} {371}},\ \bibinfo {pages} {1143} (\bibinfo {year} {2021})},\ \Eprint
  {https://arxiv.org/abs/https://www.science.org/doi/pdf/10.1126/science.abg1874}
  {https://www.science.org/doi/pdf/10.1126/science.abg1874} \BibitemShut
  {NoStop}%
\bibitem [{\citenamefont {Gao}\ \emph {et~al.}(2017)\citenamefont {Gao},
  \citenamefont {Li}, \citenamefont {Yan},\ and\ \citenamefont
  {Wang}}]{ga02017prediction}%
  \BibitemOpen
  \bibfield  {author} {\bibinfo {author} {\bibfnamefont {M.}~\bibnamefont
  {Gao}}, \bibinfo {author} {\bibfnamefont {Q.-Z.}\ \bibnamefont {Li}},
  \bibinfo {author} {\bibfnamefont {X.-W.}\ \bibnamefont {Yan}},\ and\ \bibinfo
  {author} {\bibfnamefont {J.}~\bibnamefont {Wang}},\ }\href
  {https://doi.org/10.1103/PhysRevB.95.024505} {\bibfield  {journal} {\bibinfo
  {journal} {Phys. Rev. B}\ }\textbf {\bibinfo {volume} {95}},\ \bibinfo
  {pages} {024505} (\bibinfo {year} {2017})}\BibitemShut {NoStop}%
\bibitem [{\citenamefont {Giannozzi}\ \emph {et~al.}(2009)\citenamefont
  {Giannozzi}, \citenamefont {Baroni}, \citenamefont {Bonini}, \citenamefont
  {Calandra}, \citenamefont {Car}, \citenamefont {Cavazzoni}, \citenamefont
  {Ceresoli}, \citenamefont {Chiarotti}, \citenamefont {Cococcioni},
  \citenamefont {Dabo} \emph {et~al.}}]{giannozzi2009quantum}%
  \BibitemOpen
  \bibfield  {author} {\bibinfo {author} {\bibfnamefont {P.}~\bibnamefont
  {Giannozzi}}, \bibinfo {author} {\bibfnamefont {S.}~\bibnamefont {Baroni}},
  \bibinfo {author} {\bibfnamefont {N.}~\bibnamefont {Bonini}}, \bibinfo
  {author} {\bibfnamefont {M.}~\bibnamefont {Calandra}}, \bibinfo {author}
  {\bibfnamefont {R.}~\bibnamefont {Car}}, \bibinfo {author} {\bibfnamefont
  {C.}~\bibnamefont {Cavazzoni}}, \bibinfo {author} {\bibfnamefont
  {D.}~\bibnamefont {Ceresoli}}, \bibinfo {author} {\bibfnamefont {G.~L.}\
  \bibnamefont {Chiarotti}}, \bibinfo {author} {\bibfnamefont {M.}~\bibnamefont
  {Cococcioni}}, \bibinfo {author} {\bibfnamefont {I.}~\bibnamefont {Dabo}},
  \emph {et~al.},\ }\href@noop {} {\bibfield  {journal} {\bibinfo  {journal}
  {Journal of physics: Condensed matter}\ }\textbf {\bibinfo {volume} {21}},\
  \bibinfo {pages} {395502} (\bibinfo {year} {2009})}\BibitemShut {NoStop}%
\bibitem [{\citenamefont {Giannozzi}\ \emph {et~al.}(2017)\citenamefont
  {Giannozzi}, \citenamefont {Andreussi}, \citenamefont {Brumme}, \citenamefont
  {Bunau}, \citenamefont {Nardelli}, \citenamefont {Calandra}, \citenamefont
  {Car}, \citenamefont {Cavazzoni}, \citenamefont {Ceresoli}, \citenamefont
  {Cococcioni} \emph {et~al.}}]{giannozzi2017advanced}%
  \BibitemOpen
  \bibfield  {author} {\bibinfo {author} {\bibfnamefont {P.}~\bibnamefont
  {Giannozzi}}, \bibinfo {author} {\bibfnamefont {O.}~\bibnamefont
  {Andreussi}}, \bibinfo {author} {\bibfnamefont {T.}~\bibnamefont {Brumme}},
  \bibinfo {author} {\bibfnamefont {O.}~\bibnamefont {Bunau}}, \bibinfo
  {author} {\bibfnamefont {M.~B.}\ \bibnamefont {Nardelli}}, \bibinfo {author}
  {\bibfnamefont {M.}~\bibnamefont {Calandra}}, \bibinfo {author}
  {\bibfnamefont {R.}~\bibnamefont {Car}}, \bibinfo {author} {\bibfnamefont
  {C.}~\bibnamefont {Cavazzoni}}, \bibinfo {author} {\bibfnamefont
  {D.}~\bibnamefont {Ceresoli}}, \bibinfo {author} {\bibfnamefont
  {M.}~\bibnamefont {Cococcioni}}, \emph {et~al.},\ }\href@noop {} {\bibfield
  {journal} {\bibinfo  {journal} {Journal of Physics: Condensed Matter}\
  }\textbf {\bibinfo {volume} {29}},\ \bibinfo {pages} {465901} (\bibinfo
  {year} {2017})}\BibitemShut {NoStop}%
\bibitem [{\citenamefont {Perdew}\ \emph {et~al.}(1996)\citenamefont {Perdew},
  \citenamefont {Burke},\ and\ \citenamefont
  {Ernzerhof}}]{perdew1996generalized}%
  \BibitemOpen
  \bibfield  {author} {\bibinfo {author} {\bibfnamefont {J.~P.}\ \bibnamefont
  {Perdew}}, \bibinfo {author} {\bibfnamefont {K.}~\bibnamefont {Burke}},\ and\
  \bibinfo {author} {\bibfnamefont {M.}~\bibnamefont {Ernzerhof}},\ }\href@noop
  {} {\bibfield  {journal} {\bibinfo  {journal} {Physical review letters}\
  }\textbf {\bibinfo {volume} {77}},\ \bibinfo {pages} {3865} (\bibinfo {year}
  {1996})}\BibitemShut {NoStop}%
\bibitem [{\citenamefont {Methfessel}\ and\ \citenamefont
  {Paxton}(1989)}]{PhysRevB.40.3616}%
  \BibitemOpen
  \bibfield  {author} {\bibinfo {author} {\bibfnamefont {M.}~\bibnamefont
  {Methfessel}}\ and\ \bibinfo {author} {\bibfnamefont {A.~T.}\ \bibnamefont
  {Paxton}},\ }\href {https://doi.org/10.1103/PhysRevB.40.3616} {\bibfield
  {journal} {\bibinfo  {journal} {Phys. Rev. B}\ }\textbf {\bibinfo {volume}
  {40}},\ \bibinfo {pages} {3616} (\bibinfo {year} {1989})}\BibitemShut
  {NoStop}%
\bibitem [{\citenamefont {Baroni}\ \emph {et~al.}(1987)\citenamefont {Baroni},
  \citenamefont {Giannozzi},\ and\ \citenamefont {Testa}}]{baroni1987green}%
  \BibitemOpen
  \bibfield  {author} {\bibinfo {author} {\bibfnamefont {S.}~\bibnamefont
  {Baroni}}, \bibinfo {author} {\bibfnamefont {P.}~\bibnamefont {Giannozzi}},\
  and\ \bibinfo {author} {\bibfnamefont {A.}~\bibnamefont {Testa}},\
  }\href@noop {} {\bibfield  {journal} {\bibinfo  {journal} {Physical review
  letters}\ }\textbf {\bibinfo {volume} {58}},\ \bibinfo {pages} {1861}
  (\bibinfo {year} {1987})}\BibitemShut {NoStop}%
\bibitem [{\citenamefont {Baroni}\ \emph {et~al.}(2001)\citenamefont {Baroni},
  \citenamefont {de~Gironcoli}, \citenamefont {Dal~Corso},\ and\ \citenamefont
  {Giannozzi}}]{baroni2001phonons}%
  \BibitemOpen
  \bibfield  {author} {\bibinfo {author} {\bibfnamefont {S.}~\bibnamefont
  {Baroni}}, \bibinfo {author} {\bibfnamefont {S.}~\bibnamefont
  {de~Gironcoli}}, \bibinfo {author} {\bibfnamefont {A.}~\bibnamefont
  {Dal~Corso}},\ and\ \bibinfo {author} {\bibfnamefont {P.}~\bibnamefont
  {Giannozzi}},\ }\href {https://doi.org/10.1103/RevModPhys.73.515} {\bibfield
  {journal} {\bibinfo  {journal} {Rev. Mod. Phys.}\ }\textbf {\bibinfo {volume}
  {73}},\ \bibinfo {pages} {515} (\bibinfo {year} {2001})}\BibitemShut
  {NoStop}%
\bibitem [{\citenamefont {Allen}(1972)}]{allen1972neutron}%
  \BibitemOpen
  \bibfield  {author} {\bibinfo {author} {\bibfnamefont {P.~B.}\ \bibnamefont
  {Allen}},\ }\href@noop {} {\bibfield  {journal} {\bibinfo  {journal}
  {Physical Review B}\ }\textbf {\bibinfo {volume} {6}},\ \bibinfo {pages}
  {2577} (\bibinfo {year} {1972})}\BibitemShut {NoStop}%
\bibitem [{\citenamefont {Eliashberg}(1960)}]{eliashberg1960interactions}%
  \BibitemOpen
  \bibfield  {author} {\bibinfo {author} {\bibfnamefont {G.}~\bibnamefont
  {Eliashberg}},\ }\href@noop {} {\bibfield  {journal} {\bibinfo  {journal}
  {Sov. Phys. JETP}\ }\textbf {\bibinfo {volume} {11}},\ \bibinfo {pages} {696}
  (\bibinfo {year} {1960})}\BibitemShut {NoStop}%
\bibitem [{\citenamefont {Allen}\ and\ \citenamefont
  {Mitrovi{\'c}}(1983)}]{allen1983theory}%
  \BibitemOpen
  \bibfield  {author} {\bibinfo {author} {\bibfnamefont {P.~B.}\ \bibnamefont
  {Allen}}\ and\ \bibinfo {author} {\bibfnamefont {B.}~\bibnamefont
  {Mitrovi{\'c}}},\ }in\ \href@noop {} {\emph {\bibinfo {booktitle} {Solid
  state physics}}},\ Vol.~\bibinfo {volume} {37}\ (\bibinfo  {publisher}
  {Elsevier},\ \bibinfo {year} {1983})\ pp.\ \bibinfo {pages}
  {1--92}\BibitemShut {NoStop}%
\bibitem [{\citenamefont {Allen}\ and\ \citenamefont
  {Dynes}(1975)}]{allen1975transition}%
  \BibitemOpen
  \bibfield  {author} {\bibinfo {author} {\bibfnamefont {P.~B.}\ \bibnamefont
  {Allen}}\ and\ \bibinfo {author} {\bibfnamefont {R.}~\bibnamefont {Dynes}},\
  }\href@noop {} {\bibfield  {journal} {\bibinfo  {journal} {Physical Review
  B}\ }\textbf {\bibinfo {volume} {12}},\ \bibinfo {pages} {905} (\bibinfo
  {year} {1975})}\BibitemShut {NoStop}%
\bibitem [{\citenamefont {Brumme}\ \emph {et~al.}(2014)\citenamefont {Brumme},
  \citenamefont {Calandra},\ and\ \citenamefont
  {Mauri}}]{brumme2014electrochemical}%
  \BibitemOpen
  \bibfield  {author} {\bibinfo {author} {\bibfnamefont {T.}~\bibnamefont
  {Brumme}}, \bibinfo {author} {\bibfnamefont {M.}~\bibnamefont {Calandra}},\
  and\ \bibinfo {author} {\bibfnamefont {F.}~\bibnamefont {Mauri}},\
  }\href@noop {} {\bibfield  {journal} {\bibinfo  {journal} {Physical Review
  B}\ }\textbf {\bibinfo {volume} {89}},\ \bibinfo {pages} {245406} (\bibinfo
  {year} {2014})}\BibitemShut {NoStop}%
\bibitem [{\citenamefont {Brumme}\ \emph {et~al.}(2015)\citenamefont {Brumme},
  \citenamefont {Calandra},\ and\ \citenamefont {Mauri}}]{brumme2015first}%
  \BibitemOpen
  \bibfield  {author} {\bibinfo {author} {\bibfnamefont {T.}~\bibnamefont
  {Brumme}}, \bibinfo {author} {\bibfnamefont {M.}~\bibnamefont {Calandra}},\
  and\ \bibinfo {author} {\bibfnamefont {F.}~\bibnamefont {Mauri}},\
  }\href@noop {} {\bibfield  {journal} {\bibinfo  {journal} {Physical Review
  B}\ }\textbf {\bibinfo {volume} {91}},\ \bibinfo {pages} {155436} (\bibinfo
  {year} {2015})}\BibitemShut {NoStop}%
\bibitem [{\citenamefont {Tominaka}\ \emph {et~al.}(2020)\citenamefont
  {Tominaka}, \citenamefont {Ishibiki}, \citenamefont {Fujino}, \citenamefont
  {Kawakami}, \citenamefont {Ohara}, \citenamefont {Masuda}, \citenamefont
  {Matsuda}, \citenamefont {Hosono},\ and\ \citenamefont
  {Kondo}}]{tominaka2020geometrical}%
  \BibitemOpen
  \bibfield  {author} {\bibinfo {author} {\bibfnamefont {S.}~\bibnamefont
  {Tominaka}}, \bibinfo {author} {\bibfnamefont {R.}~\bibnamefont {Ishibiki}},
  \bibinfo {author} {\bibfnamefont {A.}~\bibnamefont {Fujino}}, \bibinfo
  {author} {\bibfnamefont {K.}~\bibnamefont {Kawakami}}, \bibinfo {author}
  {\bibfnamefont {K.}~\bibnamefont {Ohara}}, \bibinfo {author} {\bibfnamefont
  {T.}~\bibnamefont {Masuda}}, \bibinfo {author} {\bibfnamefont
  {I.}~\bibnamefont {Matsuda}}, \bibinfo {author} {\bibfnamefont
  {H.}~\bibnamefont {Hosono}},\ and\ \bibinfo {author} {\bibfnamefont
  {T.}~\bibnamefont {Kondo}},\ }\href
  {https://doi.org/https://doi.org/10.1016/j.chempr.2019.11.006} {\bibfield
  {journal} {\bibinfo  {journal} {Chem}\ }\textbf {\bibinfo {volume} {6}},\
  \bibinfo {pages} {406} (\bibinfo {year} {2020})}\BibitemShut {NoStop}%
\bibitem [{\citenamefont {Mortazavi}\ \emph {et~al.}(2018)\citenamefont
  {Mortazavi}, \citenamefont {Makaremi}, \citenamefont {Shahrokhi},
  \citenamefont {Raeisi}, \citenamefont {Singh}, \citenamefont {Rabczuk},\ and\
  \citenamefont {Pereira}}]{mortazavi2018borophene}%
  \BibitemOpen
  \bibfield  {author} {\bibinfo {author} {\bibfnamefont {B.}~\bibnamefont
  {Mortazavi}}, \bibinfo {author} {\bibfnamefont {M.}~\bibnamefont {Makaremi}},
  \bibinfo {author} {\bibfnamefont {M.}~\bibnamefont {Shahrokhi}}, \bibinfo
  {author} {\bibfnamefont {M.}~\bibnamefont {Raeisi}}, \bibinfo {author}
  {\bibfnamefont {C.~V.}\ \bibnamefont {Singh}}, \bibinfo {author}
  {\bibfnamefont {T.}~\bibnamefont {Rabczuk}},\ and\ \bibinfo {author}
  {\bibfnamefont {L.~F.~C.}\ \bibnamefont {Pereira}},\ }\href
  {https://doi.org/10.1039/C7NR08725J} {\bibfield  {journal} {\bibinfo
  {journal} {Nanoscale}\ }\textbf {\bibinfo {volume} {10}},\ \bibinfo {pages}
  {3759} (\bibinfo {year} {2018})}\BibitemShut {NoStop}%
\bibitem [{\citenamefont {Katsnelson}\ and\ \citenamefont
  {Fasolino}(2013)}]{katsnelson2013graphene}%
  \BibitemOpen
  \bibfield  {author} {\bibinfo {author} {\bibfnamefont {M.~I.}\ \bibnamefont
  {Katsnelson}}\ and\ \bibinfo {author} {\bibfnamefont {A.}~\bibnamefont
  {Fasolino}},\ }\href@noop {} {\bibfield  {journal} {\bibinfo  {journal}
  {Accounts of chemical research}\ }\textbf {\bibinfo {volume} {46}},\ \bibinfo
  {pages} {97} (\bibinfo {year} {2013})}\BibitemShut {NoStop}%
\bibitem [{\citenamefont {Choi}\ \emph {et~al.}(2002)\citenamefont {Choi},
  \citenamefont {Roundy}, \citenamefont {Sun}, \citenamefont {Cohen},\ and\
  \citenamefont {Louie}}]{choi2002origin}%
  \BibitemOpen
  \bibfield  {author} {\bibinfo {author} {\bibfnamefont {H.~J.}\ \bibnamefont
  {Choi}}, \bibinfo {author} {\bibfnamefont {D.}~\bibnamefont {Roundy}},
  \bibinfo {author} {\bibfnamefont {H.}~\bibnamefont {Sun}}, \bibinfo {author}
  {\bibfnamefont {M.~L.}\ \bibnamefont {Cohen}},\ and\ \bibinfo {author}
  {\bibfnamefont {S.~G.}\ \bibnamefont {Louie}},\ }\href@noop {} {\bibfield
  {journal} {\bibinfo  {journal} {Nature}\ }\textbf {\bibinfo {volume} {418}},\
  \bibinfo {pages} {758} (\bibinfo {year} {2002})}\BibitemShut {NoStop}%
\bibitem [{\citenamefont {An}\ and\ \citenamefont
  {Pickett}(2001)}]{an2001superconductivity}%
  \BibitemOpen
  \bibfield  {author} {\bibinfo {author} {\bibfnamefont {J.~M.}\ \bibnamefont
  {An}}\ and\ \bibinfo {author} {\bibfnamefont {W.~E.}\ \bibnamefont
  {Pickett}},\ }\href {https://doi.org/10.1103/PhysRevLett.86.4366} {\bibfield
  {journal} {\bibinfo  {journal} {Phys. Rev. Lett.}\ }\textbf {\bibinfo
  {volume} {86}},\ \bibinfo {pages} {4366} (\bibinfo {year}
  {2001})}\BibitemShut {NoStop}%
\bibitem [{\citenamefont {Yildirim}\ \emph {et~al.}(2001)\citenamefont
  {Yildirim}, \citenamefont {G\"ulseren}, \citenamefont {Lynn}, \citenamefont
  {Brown}, \citenamefont {Udovic}, \citenamefont {Huang}, \citenamefont
  {Rogado}, \citenamefont {Regan}, \citenamefont {Hayward}, \citenamefont
  {Slusky}, \citenamefont {He}, \citenamefont {Haas}, \citenamefont {Khalifah},
  \citenamefont {Inumaru},\ and\ \citenamefont {Cava}}]{yildirim2001giant}%
  \BibitemOpen
  \bibfield  {author} {\bibinfo {author} {\bibfnamefont {T.}~\bibnamefont
  {Yildirim}}, \bibinfo {author} {\bibfnamefont {O.}~\bibnamefont
  {G\"ulseren}}, \bibinfo {author} {\bibfnamefont {J.~W.}\ \bibnamefont
  {Lynn}}, \bibinfo {author} {\bibfnamefont {C.~M.}\ \bibnamefont {Brown}},
  \bibinfo {author} {\bibfnamefont {T.~J.}\ \bibnamefont {Udovic}}, \bibinfo
  {author} {\bibfnamefont {Q.}~\bibnamefont {Huang}}, \bibinfo {author}
  {\bibfnamefont {N.}~\bibnamefont {Rogado}}, \bibinfo {author} {\bibfnamefont
  {K.~A.}\ \bibnamefont {Regan}}, \bibinfo {author} {\bibfnamefont {M.~A.}\
  \bibnamefont {Hayward}}, \bibinfo {author} {\bibfnamefont {J.~S.}\
  \bibnamefont {Slusky}}, \bibinfo {author} {\bibfnamefont {T.}~\bibnamefont
  {He}}, \bibinfo {author} {\bibfnamefont {M.~K.}\ \bibnamefont {Haas}},
  \bibinfo {author} {\bibfnamefont {P.}~\bibnamefont {Khalifah}}, \bibinfo
  {author} {\bibfnamefont {K.}~\bibnamefont {Inumaru}},\ and\ \bibinfo {author}
  {\bibfnamefont {R.~J.}\ \bibnamefont {Cava}},\ }\href
  {https://doi.org/10.1103/PhysRevLett.87.037001} {\bibfield  {journal}
  {\bibinfo  {journal} {Phys. Rev. Lett.}\ }\textbf {\bibinfo {volume} {87}},\
  \bibinfo {pages} {037001} (\bibinfo {year} {2001})}\BibitemShut {NoStop}%
\bibitem [{\citenamefont {d'Astuto}\ \emph {et~al.}(2007)\citenamefont
  {d'Astuto}, \citenamefont {Calandra}, \citenamefont {Reich}, \citenamefont
  {Shukla}, \citenamefont {Lazzeri}, \citenamefont {Mauri}, \citenamefont
  {Karpinski}, \citenamefont {Zhigadlo}, \citenamefont {Bossak},\ and\
  \citenamefont {Krisch}}]{astuto2007weak}%
  \BibitemOpen
  \bibfield  {author} {\bibinfo {author} {\bibfnamefont {M.}~\bibnamefont
  {d'Astuto}}, \bibinfo {author} {\bibfnamefont {M.}~\bibnamefont {Calandra}},
  \bibinfo {author} {\bibfnamefont {S.}~\bibnamefont {Reich}}, \bibinfo
  {author} {\bibfnamefont {A.}~\bibnamefont {Shukla}}, \bibinfo {author}
  {\bibfnamefont {M.}~\bibnamefont {Lazzeri}}, \bibinfo {author} {\bibfnamefont
  {F.}~\bibnamefont {Mauri}}, \bibinfo {author} {\bibfnamefont
  {J.}~\bibnamefont {Karpinski}}, \bibinfo {author} {\bibfnamefont {N.~D.}\
  \bibnamefont {Zhigadlo}}, \bibinfo {author} {\bibfnamefont {A.}~\bibnamefont
  {Bossak}},\ and\ \bibinfo {author} {\bibfnamefont {M.}~\bibnamefont
  {Krisch}},\ }\href {https://doi.org/10.1103/PhysRevB.75.174508} {\bibfield
  {journal} {\bibinfo  {journal} {Phys. Rev. B}\ }\textbf {\bibinfo {volume}
  {75}},\ \bibinfo {pages} {174508} (\bibinfo {year} {2007})}\BibitemShut
  {NoStop}%
\bibitem [{\citenamefont {Parendo}\ \emph {et~al.}(2006)\citenamefont
  {Parendo}, \citenamefont {Tan},\ and\ \citenamefont
  {Goldman}}]{parendo2006electrostatic}%
  \BibitemOpen
  \bibfield  {author} {\bibinfo {author} {\bibfnamefont {K.~A.}\ \bibnamefont
  {Parendo}}, \bibinfo {author} {\bibfnamefont {K.~S.~B.}\ \bibnamefont
  {Tan}},\ and\ \bibinfo {author} {\bibfnamefont {A.}~\bibnamefont {Goldman}},\
  }\href@noop {} {\bibfield  {journal} {\bibinfo  {journal} {Physical Review
  B}\ }\textbf {\bibinfo {volume} {73}},\ \bibinfo {pages} {174527} (\bibinfo
  {year} {2006})}\BibitemShut {NoStop}%
\bibitem [{\citenamefont {Caviglia}\ \emph {et~al.}(2008)\citenamefont
  {Caviglia}, \citenamefont {Gariglio}, \citenamefont {Reyren}, \citenamefont
  {Jaccard}, \citenamefont {Schneider}, \citenamefont {Gabay}, \citenamefont
  {Thiel}, \citenamefont {Hammerl}, \citenamefont {Mannhart},\ and\
  \citenamefont {Triscone}}]{caviglia2008electric}%
  \BibitemOpen
  \bibfield  {author} {\bibinfo {author} {\bibfnamefont {A.}~\bibnamefont
  {Caviglia}}, \bibinfo {author} {\bibfnamefont {S.}~\bibnamefont {Gariglio}},
  \bibinfo {author} {\bibfnamefont {N.}~\bibnamefont {Reyren}}, \bibinfo
  {author} {\bibfnamefont {D.}~\bibnamefont {Jaccard}}, \bibinfo {author}
  {\bibfnamefont {T.}~\bibnamefont {Schneider}}, \bibinfo {author}
  {\bibfnamefont {M.}~\bibnamefont {Gabay}}, \bibinfo {author} {\bibfnamefont
  {S.}~\bibnamefont {Thiel}}, \bibinfo {author} {\bibfnamefont
  {G.}~\bibnamefont {Hammerl}}, \bibinfo {author} {\bibfnamefont
  {J.}~\bibnamefont {Mannhart}},\ and\ \bibinfo {author} {\bibfnamefont
  {J.-M.}\ \bibnamefont {Triscone}},\ }\href@noop {} {\bibfield  {journal}
  {\bibinfo  {journal} {Nature}\ }\textbf {\bibinfo {volume} {456}},\ \bibinfo
  {pages} {624} (\bibinfo {year} {2008})}\BibitemShut {NoStop}%
\bibitem [{\citenamefont {Ahn}\ \emph {et~al.}(1999)\citenamefont {Ahn},
  \citenamefont {Gariglio}, \citenamefont {Paruch}, \citenamefont {Tybell},
  \citenamefont {Antognazza},\ and\ \citenamefont
  {Triscone}}]{ahn1999electrostatic}%
  \BibitemOpen
  \bibfield  {author} {\bibinfo {author} {\bibfnamefont {C.}~\bibnamefont
  {Ahn}}, \bibinfo {author} {\bibfnamefont {S.}~\bibnamefont {Gariglio}},
  \bibinfo {author} {\bibfnamefont {P.}~\bibnamefont {Paruch}}, \bibinfo
  {author} {\bibfnamefont {T.}~\bibnamefont {Tybell}}, \bibinfo {author}
  {\bibfnamefont {L.}~\bibnamefont {Antognazza}},\ and\ \bibinfo {author}
  {\bibfnamefont {J.-M.}\ \bibnamefont {Triscone}},\ }\href@noop {} {\bibfield
  {journal} {\bibinfo  {journal} {Science}\ }\textbf {\bibinfo {volume}
  {284}},\ \bibinfo {pages} {1152} (\bibinfo {year} {1999})}\BibitemShut
  {NoStop}%
\bibitem [{\citenamefont {Saito}\ \emph {et~al.}(2016)\citenamefont {Saito},
  \citenamefont {Nojima},\ and\ \citenamefont {Iwasa}}]{saito2016highly}%
  \BibitemOpen
  \bibfield  {author} {\bibinfo {author} {\bibfnamefont {Y.}~\bibnamefont
  {Saito}}, \bibinfo {author} {\bibfnamefont {T.}~\bibnamefont {Nojima}},\ and\
  \bibinfo {author} {\bibfnamefont {Y.}~\bibnamefont {Iwasa}},\ }\href@noop {}
  {\bibfield  {journal} {\bibinfo  {journal} {Nature Reviews Materials}\
  }\textbf {\bibinfo {volume} {2}},\ \bibinfo {pages} {1} (\bibinfo {year}
  {2016})}\BibitemShut {NoStop}%
\bibitem [{\citenamefont {Saito}\ \emph {et~al.}(2015)\citenamefont {Saito},
  \citenamefont {Kasahara}, \citenamefont {Ye}, \citenamefont {Iwasa},\ and\
  \citenamefont {Nojima}}]{saito2015metallic}%
  \BibitemOpen
  \bibfield  {author} {\bibinfo {author} {\bibfnamefont {Y.}~\bibnamefont
  {Saito}}, \bibinfo {author} {\bibfnamefont {Y.}~\bibnamefont {Kasahara}},
  \bibinfo {author} {\bibfnamefont {J.}~\bibnamefont {Ye}}, \bibinfo {author}
  {\bibfnamefont {Y.}~\bibnamefont {Iwasa}},\ and\ \bibinfo {author}
  {\bibfnamefont {T.}~\bibnamefont {Nojima}},\ }\href
  {https://doi.org/10.1126/science.1259440} {\bibfield  {journal} {\bibinfo
  {journal} {Science}\ }\textbf {\bibinfo {volume} {350}},\ \bibinfo {pages}
  {409} (\bibinfo {year} {2015})},\ \Eprint
  {https://arxiv.org/abs/https://www.science.org/doi/pdf/10.1126/science.1259440}
  {https://www.science.org/doi/pdf/10.1126/science.1259440} \BibitemShut
  {NoStop}%
\bibitem [{\citenamefont {Belli}\ \emph {et~al.}(2021)\citenamefont {Belli},
  \citenamefont {Novoa}, \citenamefont {Contreras-Garc{\'\i}a},\ and\
  \citenamefont {Errea}}]{belli2021strong}%
  \BibitemOpen
  \bibfield  {author} {\bibinfo {author} {\bibfnamefont {F.}~\bibnamefont
  {Belli}}, \bibinfo {author} {\bibfnamefont {T.}~\bibnamefont {Novoa}},
  \bibinfo {author} {\bibfnamefont {J.}~\bibnamefont {Contreras-Garc{\'\i}a}},\
  and\ \bibinfo {author} {\bibfnamefont {I.}~\bibnamefont {Errea}},\
  }\href@noop {} {\bibfield  {journal} {\bibinfo  {journal} {Nature
  Communications}\ }\textbf {\bibinfo {volume} {12}},\ \bibinfo {pages} {1}
  (\bibinfo {year} {2021})}\BibitemShut {NoStop}%
\end{thebibliography}%

\end{document}